**Experimental and computational diffusion analysis in Ni-X binary and Ni-Al-X (X = Cr, Mo, Ta, W, Re) ternary systems**

Ankur Srivastava[1], Suman Sadhu[1], Satyam Kumar[1], Ujjval Bansal[1], Raju Ravi[1], Saswata Bhattacharyya[2], Gopalakrishnan Sai Gautam[1], Aloke Paul[1,*]

[1]Department of Materials Engineering, Indian Institute of Science, Bengaluru, Karnataka, 560012, India

[2]Department of Material Science and Metallurgical Engineering, Indian Institute of Technology Hyderabad, Telangana, 502284, India

*Corresponding authors: sankur@iisc.ac.in, saswata@msme.iith.ac.in, saigautamg@iisc.ac.in, aloke@iisc.ac.in

## Abstract

An extensive diffusion analysis is presented for binary Ni-X and ternary Ni-Al-X (X = Cr, Mo, Ta, W, Re) systems, which play a crucial role in microstructural evolution and phase stability in Ni-Al-based superalloys. Specifically, we highlight changes in the diffusion coefficients of X in the presence of Al and compare diffusional interactions across systems considered. First-principles calculations, combined with activation energies derived from temperature-dependent experiments, reveal consistent trends in Ni-X systems, with variations in activation energies largely attributed to differences in migration energies. In ternary systems, diffusion coefficients estimated from intersecting diffusion profiles show that the main interdiffusion coefficient of X is comparable to its binary counterpart, with similar activation energies. However, cross-diffusion coefficients are shown to significantly influence fluxes, either enhancing or reducing diffusion lengths depending on the relative directions of diffusing elements. For Ni-Al-Re, a single-profile method is employed to overcome uncertainties in estimating composition gradients at the near-end-member intersecting composition. The diffusion coefficients obtained correlate well with the nature of diffusion paths when represented on Gibbs triangles. To extend these findings, a physics-informed neural network (PINN) optimization method is applied to extract composition-dependent diffusion coefficients across the full composition range. The analysis demonstrates the necessity of incorporating experimentally estimated diffusion coefficients as equality constraints, without which optimization reliability is compromised. Overall, the results establish a robust framework for diffusion studies in Ni-Al-X systems, highlighting the critical role of cross-diffusion effects and constraint-enhanced numerical methods.

**Keywords:** Diffusion; Ni-based alloys; First Principles calculations; PINN optimisation

## 1. Introduction

Diffusion studies in Ni-based superalloys are crucial as the microstructure evolution, high-temperature material properties, and service life at high-temperature applications depend on diffusion-controlled processes. The elements, for example, X = Cr, Mo, Ta, W, Re, are the primary alloying additions in Ni-Al-based superalloys among VB to VIIB elements in the periodic table, which are crucial for developing desired microstructure, achieving the required microstructural stability, and providing high-temperature mechanical properties and oxidation resistance.

Most diffusion studies to date have been conducted primarily in binary systems, Ni-X [1-17], and Ni-Al [18,19], with very limited studies in ternary Ni-Al-X systems [20-27]. The diffusion coefficients reported in the literature for binary systems show a significant difference in activation energy within a given system. Moreover, studies of binary Ni-X systems do not indicate diffusional interactions (as measured by cross-diffusion coefficients) in the presence of Al. This is important, as the flux of an element may increase or decrease depending on the sign of the cross-diffusion coefficients and diffusion directions of the elements.

Diffusion studies in ternary Ni-Al-X systems are mostly conducted at specific temperatures, without reporting activation energies, except for Ni-Al-Re [20-27]. A maximum number of studies are available on Ni-Al-Re solid solutions [24-27], as Re plays a crucial role in microstructural stabilisation and creep resistance due to its very low diffusion coefficients. Similar to the binary Ni-Re system, the activation energies in the Ni-Al-Re system are also reported to vary significantly across studies. In all the systems, only the estimated interdiffusion coefficients ($\widetilde{D}_{ij}^{n}$) are reported, which are a kind of average of the intrinsic ($D_{ij}^{n}$) and tracer ($D_i^*$) diffusion coefficients of elements. Calculation of $D_{ij}^{n}$ and $D_i^*$ are important to understand the diffusion rates of an element in the presence and absence of a thermodynamic driving force.

The interdiffusion rates are reported at specific temperatures without a comparison across different systems. Moreover, the change in diffusion rate of Al in the presence of these elements is yet to be addressed. Since diffusion coefficients are experimentally estimated at a particular composition, a numerical inverse method is essential for extracting composition-dependent diffusion coefficients over the entire composition range of a diffusion couple. A type of numerical inverse method was previously utilised in Ni-Al-Ta [22] and Ni-Al-Re [26] systems; however, we have recently shown that single-profile optimisation does not guarantee a reliable outcome unless experimentally estimated diffusion coefficients are used as equality constraints at multiple compositions [28]. Additionally, the match in $\widetilde{D}_{ij}^{n}$ at a particular composition does not guarantee reliable output of $D_{ij}^{n}$ and $D_{i}^{*}$ unless optimized with these experimentally estimated data as the equality constraints.

The primary objective of this study is to investigate diffusion phenomena in ternary Ni–Al–X systems and address the challenges outlined above. To establish a comparative framework, an in-depth analysis of binary systems is first undertaken, given the difficulty of identifying reliable diffusion coefficients from scattered datasets and the limited availability of activation energies in the literature. Complementing this, first-principles calculations of activation energies for impurity diffusion are performed to elucidate system-specific differences by comparing the vacancy formation and migration energies of X with experimentally estimated activation energies, which cannot distinguish the specific roles of these factors.

Building on this foundation, all types of diffusion coefficients ($D_{i}^{*}$, $D_{ij}^{n}$, and $\widetilde{D}_{ij}^{n}$) are systematically estimated to capture distinct aspects of the diffusion process. This comprehensive treatment underscores why interpretations based exclusively on $\widetilde{D}_{ij}^{n}$ can lead to misleading conclusions regarding diffusional interactions. The estimation of cross-diffusion coefficients reveals how the simultaneous presence of Al and X modifies diffusion behavior, with coefficients either reduced or enhanced depending on the sign of the cross terms and the relative diffusion directions of the elements.

For most systems, these coefficients are derived by intersecting two diffusion paths. However, in the Ni–Al–Re system, path intersections occurred near end-member compositions, introducing uncertainty in the determination of composition gradients. To overcome this, a recently proposed method for estimating the Kirkendall marker plane from a single diffusion profile [29] was employed. Subsequently, a constraint-enhanced physics-informed neural network (PINN) approach was applied to extract diffusion coefficients across the full composition range. This analysis demonstrates the necessity of experimentally determined diffusion coefficients as constraints, without which reliable optimization parameters cannot be achieved. The extendibility of optimized parameters developed in ternary systems is validated through comparison with both calculated and experimentally estimated $\widetilde{D}$ values in binary systems, thereby reinforcing the methodological robustness and transferability of the present framework.

## 2. Experimental and Computational methods

### *2.1 Diffusion couple experiments*

End-member alloy buttons were synthesized by arc melting mixtures of high-purity elements (99.95–99.99 wt.%) under an argon atmosphere. The chamber was first evacuated to ~$10^{-4}$ Pa and subsequently backfilled with Ar, following which each alloy button was remelted five to six times to ensure thorough homogenization. Alloys not containing rhenium were subjected to homogenization treatment at 1200 °C for 50 h in a vacuum tube furnace (~$10^{-4}$ Pa), whereas Re-containing alloys were homogenized at 1250 °C for 50 h because of Re's comparatively lower diffusion coefficient.

The average compositions of the homogenized alloy buttons were determined at randomly selected positions after metallographic preparation, using wavelength-dispersive spectroscopy (WDS) in an electron-probe micro-analyser (EPMA). Pure elements were used as calibration standards. Target and measured compositions of alloys prepared for binary and ternary diffusion couples are presented in Table 1.

Thin slices (~1.5 mm) were sectioned from the homogenized alloy buttons using electro-discharge machining. These slices were assembled into diffusion couples within a fixture in the vacuum tube furnace for annealing. Binary diffusion couples were annealed at four temperatures between 1100 and 1250 °C for 50 h. Ternary Ni–Al–X diffusion couples (X = Cr, Mo, Ta, W, Re) were annealed at three temperatures in the range 1100–1200 °C. Re-containing ternary couples were annealed at three temperatures between 1150 and 1250 °C for 50 h.

Upon completion of annealing, the diffusion couples were cross-sectioned using a slow-speed diamond saw and prepared metallographically for WDS line-profile analysis in the EPMA.

| Binary Systems | | | | |
|---|---|---|---|---|
| Ni-X systems | Diffusion Couple | End Member | Target Composition | Actual Composition |
| Ni-Cr | B1 | EM1 | Ni | Ni |
| | | EM2 | $Ni_{95}Cr_{5}$ | $Ni_{94.8}Cr_{5.2}$ |
| Ni-Mo | B2 | EM1 | Ni | Ni |
| | | EM2 | $Ni_{95}Mo_{5}$ | $Ni_{95.2}Mo_{4.8}$ |
| Ni-Ta | B3 | EM1 | Ni | Ni |
| | | EM2 | $Ni_{95}Ta_{5}$ | $Ni_{94.8}Ta_{5.2}$ |
| Ni-W | B4 | EM1 | Ni | Ni |
| | | EM2 | $Ni_{95}W_{5}$ | $Ni_{94.8}W_{5.2}$ |
| Ni-Re | B5 | EM1 | Ni | Ni |
| | | EM2 | $Ni_{95}Re_{5}$ | $Ni_{95.4}Re_{4.6}$ |
| Ternary Systems | | | | |
| Ni-Al-X systems | Diffusion Couple | End Member | Target Composition | Actual Composition |
| Ni-Al-Cr | T1 | EM1 | $Ni_{95}Cr_{5}$ | $Ni_{94.9}Cr_{5.1}$ |
| | | EM2 | $Ni_{95}Al_{5}$ | $Ni_{94.6}Al_{5.4}$ |
| | T2 | EM1 | Ni | Ni |
| | | EM2 | $Ni_{90}Cr_{5}Al_{5}$ | $Ni_{89.4}Cr_{5.1}Al_{5.5}$ |
| Ni-Al-Mo | T3 | EM1 | $Ni_{95}Mo_{5}$ | $Ni_{95.2}Mo_{4.7}$ |
| | | EM2 | $Ni_{95}Al_{5}$ | $Ni_{94.7}Al_{5.3}$ |
| | T4 | EM1 | Ni | Ni |
| | | EM2 | $Ni_{90}Mo_{5}Al_{5}$ | $Ni_{89.8}Mo_{4.7}Al_{5.5}$ |
| Ni-Al-Ta | T5 | EM1 | $Ni_{95}Ta_{5}$ | $Ni_{95.3}Ta_{4.7}$ |
| | | EM2 | $Ni_{95}Al_{5}$ | $Ni_{94.7}Al_{5.3}$ |
| | T6 | EM1 | Ni | Ni |
| | | EM2 | $Ni_{90}Ta_{5}Al_{5}$ | $Ni_{89.8}Ta_{4.9}Al_{5.3}$ |
| Ni-Al-W | T7 | EM1 | $Ni_{95}W_{5}$ | $Ni_{95.6}W_{4.3}$ |
| | | EM2 | $Ni_{95}Al_{5}$ | $Ni_{94.7}Al_{5.3}$ |
| | T8 | EM1 | Ni | Ni |
| | | EM2 | $Ni_{90}W_{5}Al_{5}$ | $Ni_{89.8}W_{4.9}Al_{5.3}$ |
| Ni-Al-Re | T9 | EM1 | $Ni_{95}Re_{5}$ | $Ni_{95.5}Re_{4.5}$ |
| | | EM2 | $Ni_{95}Al_{5}$ | $Ni_{94.7}Al_{5.3}$ |

Table 1: Alloy target and actual compositions used to produce the binary and ternary diffusion couples. B1-5 represents binary diffusion couples, and T1-9 represents ternary diffusion couples. EM1 and EM2 represent end-member alloys of a particular diffusion couple.

*2.2 First Principles Calculations*

We evaluated the activation energies for the diffusion of different solute atoms to the nearest vacancy in Ni (i.e., the impurity diffusion coefficient) using first-principles calculations [30,31]. We computed defect formation and transition-state (migration) energies using spin-polarised density functional theory (DFT) with projector augmented wave [32] potentials, as implemented in the Vienna ab initio simulation package [33,34]. We described the electronic exchange-correlation interactions using the generalised gradient approximation (GGA), as parameterised by Perdew, Burke, and Ernzerhof [35]. The irreducible Brillouin zone for all systems was sampled using Γ-centered $k$-meshes generated with the Monkhorst-Pack scheme [36] at a density of at least 48/Å. We expanded the one-electron wavefunctions in a plane-wave basis set with a kinetic-energy cutoff of 520 eV. We relaxed the cell volume, cell shape, and atomic positions for all input structures until the atomic forces and total energies converged below the value of $|0.01|$ eV/Å and $10^{-5}$ eV, respectively.

The formation energy of a vacancy $E_V^X$ (eV) is given by

$$E_V^X = E_{vac} - E_{bulk} - \mu_i N_i \quad (1)$$

where $E_{vac}$ and $E_{bulk}$ are the total energies of a supercell with and without a vacancy, respectively. $n_i$ represents the number of atoms of Ni removed (*i.e.* one) to form the vacancy, with $\mu_i$ the corresponding chemical potential of the Ni. We referenced the $\mu_i$ to the GGA-calculated energy of Ni in its ground state face-centred-cubic (FCC) structure at 0 K. For all defective structures, we relaxed only the atomic positions while keeping the lattice parameters to those of the relaxed bulk (i.e., non-defective) structure. For all defect calculations, we used a 3×3×3 supercells of the FCC unit cell, totalling 108 atoms.

The minimum-energy paths and migration barriers for vacancy-mediated jumps were evaluated using the nudged elastic band (NEB) method [37,38] in conjunction with DFT. A force convergence criterion of |0.05| eV/Å was used in our NEB calculations for components perpendicular to the migration path tangent. Seven images, equally spaced between the endpoints as an initial guess, were considered with a spring force constant of 5 eV/Å$^2$ between images. We performed all NEB calculations in 3×3×3 supercells of the FCC unit cell with one Ni atom replaced by a solute atom (i.e., X = Cr, Mo, Ta, Re, W) to ensure a minimum of ~ 8 Å distance between periodic images, thus reducing fictitious image–image interactions. We used a $k$-point mesh density of 48/Å for relaxing the endpoints and 32/Å for the actual NEB calculation to reduce computational costs.

*2.3 Physics-informed neural network (PINN) numerical optimisation method*

The tracer diffusion coefficient ($D_i^*$) is composition-dependent; for a consistent non-dimensional formulation and improved numerical stability during optimisation, we introduce its dimensionless form $\overline{D}_i^*$ [28]:

$$\overline{D}_i^* = \frac{D_i^*}{D_0} \tag{2}$$

where $D_0 = \frac{L_n^2}{T_n}$ $(m^2/s)$ is a characteristic diffusivity scale defined using a reference length $L_n$ $(m)$ and a reference time $T_n$ $(s)$.

The composition dependence of this dimensionless $D_i^*$ for the ternary Ni-Al-X equivalent to system is expressed as:

$$\ln(\overline{D}_i^*) = \theta_0^i + \theta_i^{1,i} N_i + \theta_j^{1,i} N_j + \theta_{i,j}^{2,i} N_i N_j \tag{3}$$

where $\Theta = \{\theta_0^i, \theta_i^{1,i}, \theta_j^{1,i}, \theta_{i,j}^{2,i}\}$ denotes the trainable parameter set. Here $N_i$ and $N_j$ (corresponding to $N_{Al}$ and $N_X$) are the independent mole fractions, while $N_{Ni}$ is the dependent mole fraction obtained from the constraint $\sum N_i = 1$. Since $\overline{D}_i^*$ is dimensionless, the coefficients appearing in Eq. 3 are also dimensionless. However, this logarithmic representation may be interpreted in analogy with an Arrhenius-type expression, $D_i^* = D_i^0 \exp\left(-\frac{Q_i}{RT}\right)$, where $D_i^0$is the pre-exponential factor, $Q_i$is an effective activation energy, $R$ is the universal gas constant, and $T$is the absolute temperature; in this form, the composition dependence is carried through an effective activation term.

In this interpretation, the coefficients in Eq. 3 may be viewed as activation-energy-like contributions normalized by $RT$. Specifically, if $\Theta_0^i, \Theta_i^{1,i}, \Theta_j^{1,i}, \Theta_{i,j}^{2,i}$ denote the corresponding dimensional energetic parameters (with units of J/mol), then the trainable coefficients are their dimensionless scaled forms, given by:

$$\theta_0^i = \frac{\Theta_0^i}{RT}, \theta_i^{1,i} = \frac{\Theta_i^{1,i}}{RT}, \theta_j^{1,i} = \frac{\Theta_j^{1,i}}{RT}, \theta_{i,j}^{2,i} = \frac{\Theta_{i,j}^{2,i}}{RT} \tag{3.1}$$

The constant term additionally absorbs the contribution from the pre-factor ratio $D_i^0/D_0$. This nondimensional logarithmic form improves numerical conditioning and numerical stability during optimisation and is convenient for implementation in DeepXDE [39].

Following the Onsager theorem, the intrinsic flux of an element $J_i$ of species $i$ can be written as [40-44]:

$$J_i = -\sum_{j=1}^{n} L_{ij} \frac{\partial \mu_j}{\partial x}, \tag{4}$$

where $L_{ij}$ are the Onsager phenomenological coefficients and $\partial\mu_j/\partial x$ is the chemical potential gradient of species $j$. Manning converted this correlation to express it with the $D_i^*$ by establishing the correlations as [44].

$$L_{ii} = \frac{C_i D_i^*}{RT}\left(1 + \xi N_i D_i^*\right) \qquad L_{ij} = \frac{C_j D_j^*}{RT}\left(\xi N_i D_i^*\right), \quad (i \neq j) \tag{5}$$

where $\xi = \frac{2}{M_0 \sum_m N_m D_m^*}$ and $M_0$ is a constant related to the crystal structure (7.15 for the FCC phase considered in this study). Assuming a constant molar volume $V_m$, the concentration of species $i$ is written as $C_i = N_i/V_m$; accordingly, substituting Eq. 5 into Eq. 4 and multiplying throughout by $V_m$ yields Eq. 6:

$$V_m J_i = -\frac{N_i D_i^*}{RT}\frac{\partial \mu_i}{\partial x} - \frac{\xi N_i D_i^*}{RT}\sum_{j=1}^{n} N_j D_j^* \frac{\partial \mu_j}{\partial x} \tag{6}$$

Further, it was shown that

$$\frac{\partial \mu_i}{\partial x} = \frac{RT}{N_1}\phi_{i1}\frac{\partial N_1}{\partial x} + \frac{RT}{N_2}\phi_{i2}\frac{\partial N_2}{\partial x} + \cdots + \frac{RT}{N_n}\phi_{in}\frac{\partial N_n}{\partial x} = RT\sum_{j=1}^{n}\frac{\phi_{ij}}{N_j}\frac{\partial N_j}{\partial x} \tag{7a}$$

Taking element n as dependent, we can further write the Eq. 7a as,

$$\begin{aligned}\frac{\partial \mu_i}{\partial x} &= \frac{RT}{N_1}\left(\phi_{i1} - \frac{N_1}{N_n}\phi_{in}\right)\frac{\partial N_1}{\partial x} + \frac{RT}{N_2}\left(\phi_{i2} - \frac{N_2}{N_n}\phi_{in}\right)\frac{\partial N_2}{\partial x} + \cdots \\ &= \frac{RT}{N_1}\phi_{i1}^n\frac{\partial N_1}{\partial x} + \frac{RT}{N_2}\phi_{i2}^n\frac{\partial N_2}{\partial x} + \cdots \ldots + \frac{RT}{N_{n-1}}\phi_{i(n-1)}^n\frac{\partial N_{n-1}}{\partial x} = RT\sum_{j=1}^{n-1}\frac{\phi_{i1}^n}{N_j}\frac{\partial N_j}{\partial x}\end{aligned} \tag{7b}$$

where $\phi_{ij}$ is the thermodynamic factor relating the activity of a species $i$ ($a_i$) to the composition of species $j$, defined as $\phi_{ij} = \left(\frac{\partial ln a_i}{\partial ln N_j}\right)_{p,T}$ and $\phi_{ij}^n = \left(\phi_{ij} - \frac{N_j}{N_n}\phi_{in}\right)$.

If the vacancy wind effect is neglected, the $\xi$-dependent second term in Eq. 6 vanishes, and only the direct $D^*$ contribution is retained. Thus, after substituting Eq. 7a in Eq. 6, it reduces to

$$V_m J_i = -\frac{N_i D_i^*}{RT} \sum_{j=1}^{n} \frac{RT}{N_j} \ \phi_{ij} \frac{\partial N_j}{\partial x} \tag{8a}$$

or Substituting Eq. 7b in Eq. 6,

$$V_m J_i = -\frac{N_i D_i^*}{RT} \sum_{j=1}^{n-1} \frac{RT}{N_j} \phi_{ij}^{n} \frac{\partial N_j}{\partial x} \tag{8b}$$

The interdiffusion flux $\tilde{J}_i$ can be directly related to the intrinsic flux following [40,43]:

$$\tilde{J}_i = J_i - N_i \sum_{k=1}^{n} J_k \tag{9}$$

Therefore. Substituting Eq. 8 in Eq. 9, we can express the interdiffusion flux directly with the $D_i^*$.

For optimisation in terms of the $\widetilde{D}_{ij}^{n}$, the composition dependence is represented using the same functional form as in Eq. 3. For simplicity, all $\widetilde{D}_{ij}^{n}$ are parameterized using a common trainable parameter set, such that

$$ln\left(\overline{\widetilde{D}}_{ij}^{n}\right) = \theta_0^i + \theta_i^{1,i} N_i + \theta_j^{1,i} N_j + \theta_{i,j}^{2,i} N_i N_j \tag{10}$$

The interdiffusion fluxes are related to the $\widetilde{D}$ following

$$V_m \tilde{J}_i = -\sum_{j=1}^{n-1} \widetilde{D}_{ij}^{n} \frac{dN_j}{dx} \tag{11a}$$

In a ternary system, we have

$$V_m \tilde{J}_i = -\widetilde{D}_{ii}^{n} \frac{dN_i}{dx} - \widetilde{D}_{ij}^{n} \frac{dN_j}{dx}, \tag{11b}$$

$$V_m \tilde{J}_j = -\widetilde{D}_{ji}^{n} \frac{dN_i}{dx} - \widetilde{D}_{jj}^{n} \frac{dN_j}{dx}, \tag{11c}$$

$$\tilde{J}_i + \tilde{J}_j + \tilde{J}_n = 0 \tag{11d}$$

where $V_m$ is the constant molar volume, $\widetilde{D}_{ii}^{n}$ and $\widetilde{D}_{jj}^{n}$ are the main interdiffusion coefficients, $\widetilde{D}_{ij}^{n}$ and $\widetilde{D}_{ji}^{n}$ are the cross-interdiffusion coefficients. Element $n$ is considered the dependent variable. The interdiffusion fluxes can be directly calculated from the diffusion profiles following

$$\tilde{J}_i = -\frac{1}{2t}\left[(1 - Y_i^*) \int_{x^{-\infty}}^{x^*} Y_i dx + Y_i^* \int_{x^*}^{x^{+\infty}} (1 - Y_i) dx\right] \tag{12}$$

where $Y_i = \frac{N_i^* - N_i^-}{N_i^+ - N_i^-}$ is the Sauer-Freise composition normalisation variable [45]. $N_i^-$ and $N_i^+$ are the compositions of the unaffected left and right-hand ends of the diffusion couple, $N_i^*$ is the composition of interest at the position $x^*$, $t$ is the annealing time.

The time-dependent evolution of the diffusion profile is related to Fick's second law for constant molar volume as

$$\frac{\partial N_i}{\partial t} = -\frac{\partial (V_m \tilde{J}_i)}{\partial x}, \tag{13}$$

where $x$is the spatial coordinate along the diffusion couple, and $t$ is the annealing time. The diffusion profile is optimised after a certain annealing time at the temperature of interest. Therefore, utilising the Boltzmann parameter, $\lambda = \frac{x}{\sqrt{t}}$, we can rewrite Eq. 13 as an ordinary differential equation (ODE) as [28]

$$-\frac{\lambda}{2}\frac{dN_i}{d\lambda} + \frac{d[V_m \tilde{J}_i(\lambda)]}{d\lambda} = 0 \tag{14}$$

For numerical convenience, $\lambda$ is normalised to $\bar{\lambda} \in [0,1]$ following

$\bar{\lambda} = \frac{\lambda - \lambda^-}{\lambda^+ - \lambda^-}$, where $\lambda^-$ and $\lambda^+$ are the Boltzmann parameters corresponding to the unaffected left and right end-member regions, respectively. Defining $\Delta\lambda = \lambda^+ - \lambda^-$, it follows that

$\lambda = \lambda^- + \Delta\lambda\,\bar{\lambda}$ and $d\lambda = \Delta\lambda d\bar{\lambda}$. Using these relations, Eq. 14 can be rewritten as

$$-\frac{(\lambda^- + \Delta\lambda\,\bar{\lambda})}{2}\frac{dN_i}{d\bar{\lambda}} + \frac{d}{\Delta\lambda d\bar{\lambda}}\left[V_m \tilde{J}_i(\bar{\lambda})\right] = 0, \tag{15}$$

This normalized ODE is used as the governing physics constraint in the PINN optimisation. Since diffusion-couple data are available after a prescribed annealing time, this formulation is well-suited for inverse modelling and significantly reduces the computational cost.

The inverse problem is posed by seeking a composition profile $N_i(\bar{\lambda})$and a parameter set $\Theta$that satisfy the governing physics while matching both the measured diffusion profile and the diffusivity constraints. In compact form, the problem consists of the ODE residual

$-\frac{(\lambda^- + \Delta\lambda \times \bar{\lambda})}{2}\frac{dN_i}{\Delta\lambda d\bar{\lambda}} + \frac{d}{d\bar{\lambda}}\left[V_m \tilde{J}_i(\bar{\lambda})\right] = 0$, the boundary conditions: $N_i(0) = N_i^L, N_i(1) = N_i^R$, and equality constraints imposed at selected compositions $D_i^*(N_u) = D_i^{\text{eq}}(N_u)$. All equality constraints are enforced in the same scaled form, replacing $D_i^*$ by $D_0 \bar{D}_i^*$ and $D_i^{\text{eq}}$ by $\bar{D}_i^{\text{eq}} =$ $D_i^{\mathrm{eq}}/D_0$. This rescaling leaves the governing physics unchanged while improving the numerical balance of the loss terms.

Experimentally estimated diffusion coefficients are incorporated as additional constraints. The total loss function is written as

$$\mathcal{L}_{\mathrm{total}} = \omega_s \mathcal{L}_s + \omega_b \mathcal{L}_b + \omega_d \mathcal{L}_d + \omega_c \mathcal{L}_c \ , \tag{16}$$

where $\mathcal{L}_s, \mathcal{L}_b, \mathcal{L}_d$ and $L_c$ denote the ODE residual loss, boundary-condition loss, data-fitting loss, and auxiliary constraint loss, respectively. Using the normalized governing equation, the ODE residual loss is defined as $\mathcal{L}_s = \frac{1}{P_s}\sum_{p=1}^{P_s}\left[-\frac{(\lambda^- + \Delta\lambda\times\bar{\lambda})}{2}\frac{dN_i}{d\bar{\lambda}} + \frac{d}{\Delta\lambda d\bar{\lambda}}\left[V_m\tilde{J}_i(\bar{\lambda})\right]\right]^2$. The boundary-condition loss is $\mathcal{L}_b = \frac{1}{P_b}\sum_{p=1}^{P_b}[(N_i(0) - N_i^L)^2 + (N_i(1) - N_i^R)^2]$. The data-fitting loss is written as $\mathcal{L}_d = \sum_{p=1}^{P_d}\frac{1}{P_d}\left|\bar{N}_i(\bar{\lambda}_p) - N_i^p(\theta_j)\right|^2$, and the auxiliary constraint loss is $\mathcal{L}_c = \frac{1}{P_c}\sum_{p=1}^{P_c}\left|\bar{w}(\bar{\lambda}_p) - w_p(\Theta)\right|^2$.

Here $P_s, P_b, P_d, P_c$ denote the numbers of collocation points used for the ODE residual, boundary-condition points, experimental data points, and auxiliary constraint points, respectively. $\bar{N}_i(\bar{\lambda}_p)$ and $N_i^p(\theta_j)$ are the experimental and predicted compositions at $\bar{\lambda}_p$, respectively. Similarly, $\bar{w}(\bar{\lambda}_p)$ and $w_p(\Theta)$ denote the measured (or estimated) and predicted tracer or impurity diffusivities at selected locations, such as the intersecting composition of diffusion profiles or the Kirkendall marker plane. The weights $\omega_s, \omega_b, \omega_d, \omega_c$ determine the relative contributions of the corresponding loss terms [28].

Gradients are computed using automatic differentiation, which accounts for both the explicit dependence of the loss on the trainable parameters and the implicit dependence through the predicted composition field $N_i$. Writing the ODE constraint abstractly as $j(\Theta, N_i) = 0$, the total derivative of the objective with respect to $\theta_j \in \Theta$ may be expressed as

$$\frac{d\hat{L}(\Theta)}{d\theta_j} = \nabla_{\theta_j} L(\Theta, N_i(\Theta)) + \nabla_{N_i} L(\Theta, N_i(\Theta))\frac{dN_i(\Theta)}{d\theta_j}, j = 1,2, .... \tag{17a}$$

while the associated constraint sensitivity satisfies

$$\nabla_{\theta_j} j + \nabla_{N_i} j \frac{dN_i(\Theta)}{d\theta_j} = 0. \tag{17b}$$

In practice, training is performed using ADAM with a learning rate warm-up and decay [46]. The same fixed reference diffusivity, $D_0 = \frac{L_n^2}{T_n} = \frac{(1\mu m)^2}{1\,hr} = 2.77 \times 10^{-16}\,\frac{m^2}{s}$, is used throughout each training run.

**3. Results and discussion**

The objective of this study is to perform a comprehensive diffusion analysis of Ni–Al–X systems (X = Cr, Mo, Ta, W, Re), with particular emphasis on diffusional interactions among the constituent elements. To establish a comparative framework, preliminary experiments are first undertaken in the binary Ni–X systems. These enable the assessment of the influence of adding Al as a third component.

*3.1 Diffusion analysis in the binary system*

*3.1.1 Experimental estimation of diffusion coefficient*

Binary diffusion couples were prepared by joining pure Ni with $Ni_{95}X_5$ alloys and annealing them in the temperature range of 1050–1200 °C for 50 h (see Table 1). Fig. 1 illustrates the diffusion couples obtained at 1200 °C across all Ni–X binary systems. Analysis of the diffusion profiles reveals that Cr and Ta exhibit comparable diffusion lengths, while Mo shows a slightly reduced value. In contrast, the diffusion lengths of W and Re are significantly lower, with Re displaying the shortest diffusion length among the investigated elements. These observations reflect the relative diffusion coefficients of the alloying elements in Ni.

The binary interdiffusion coefficients $(\widetilde{D})$ can be calculated from the Sauer-Freise relation expressed for constant molar volume as [45, 47]

$$\widetilde{D} = \frac{1}{2t}\left(\frac{\partial x}{\partial Y}\right)_{x^*}\left[(1 - Y_i^*)\int_{x^{-\infty}}^{x^*} Y_i dx + Y_i^* \int_{x^*}^{x^{+\infty}} (1 - Y_i) dx\right] \quad (18)$$

The composition-dependent estimates of $\widetilde{D}$ are presented in Fig. 2. Values reported at 2.5 at.% from various literature sources are compared in Table 2, revealing both agreement and occasional discrepancies. Interdiffusion analysis for the Ni–Al binary system was not undertaken in the present study, since closely matching data are already available in Refs. [18,19] (see Table 2)

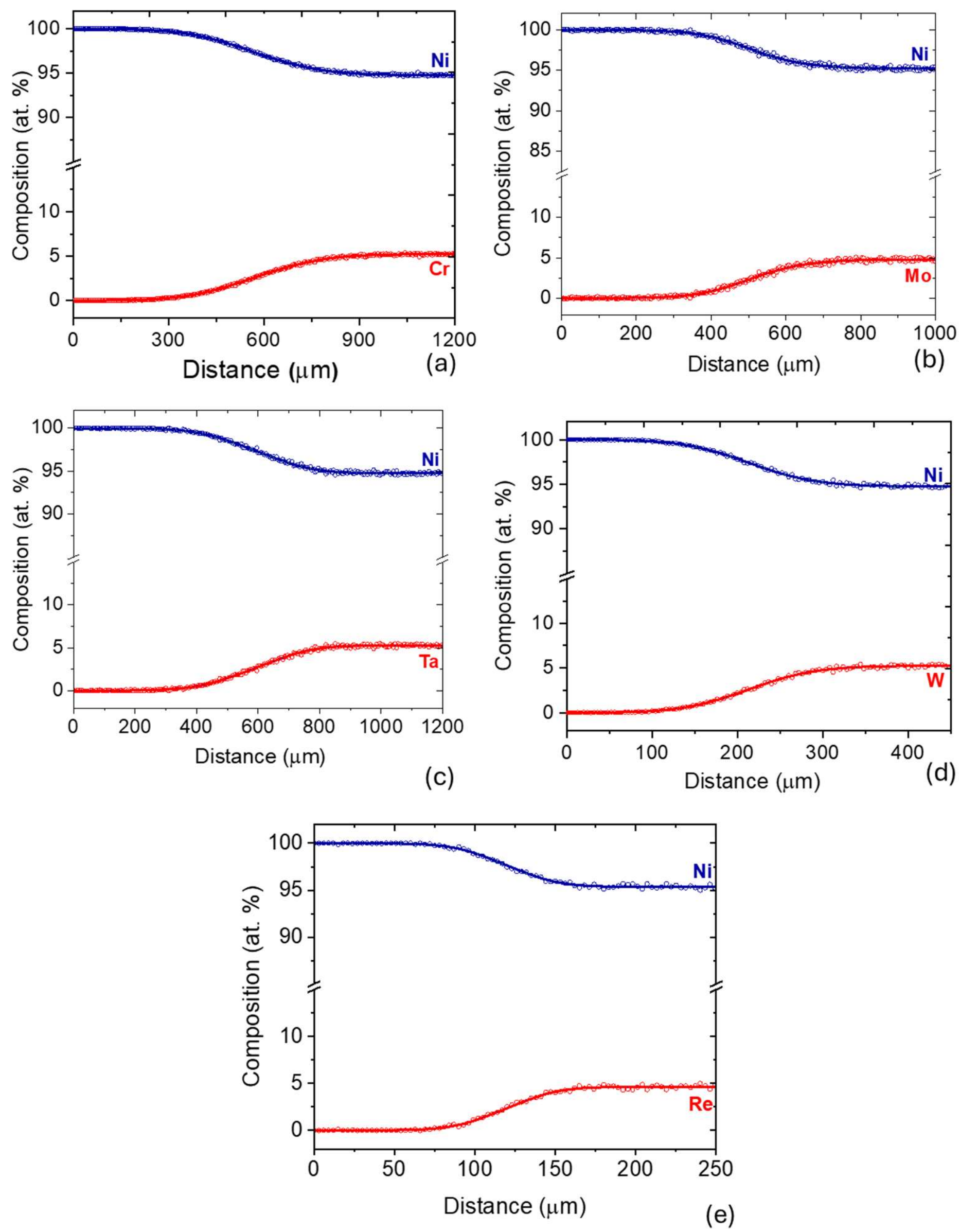

Fig. 1 Diffusion profiles developed in Ni-X systems at 1200 ℃ after annealing for 50 h (a) Ni-Cr, (b) Ni-Mo, (c) Ni-Ta, (d) Ni-W and (e) Ni-Re.

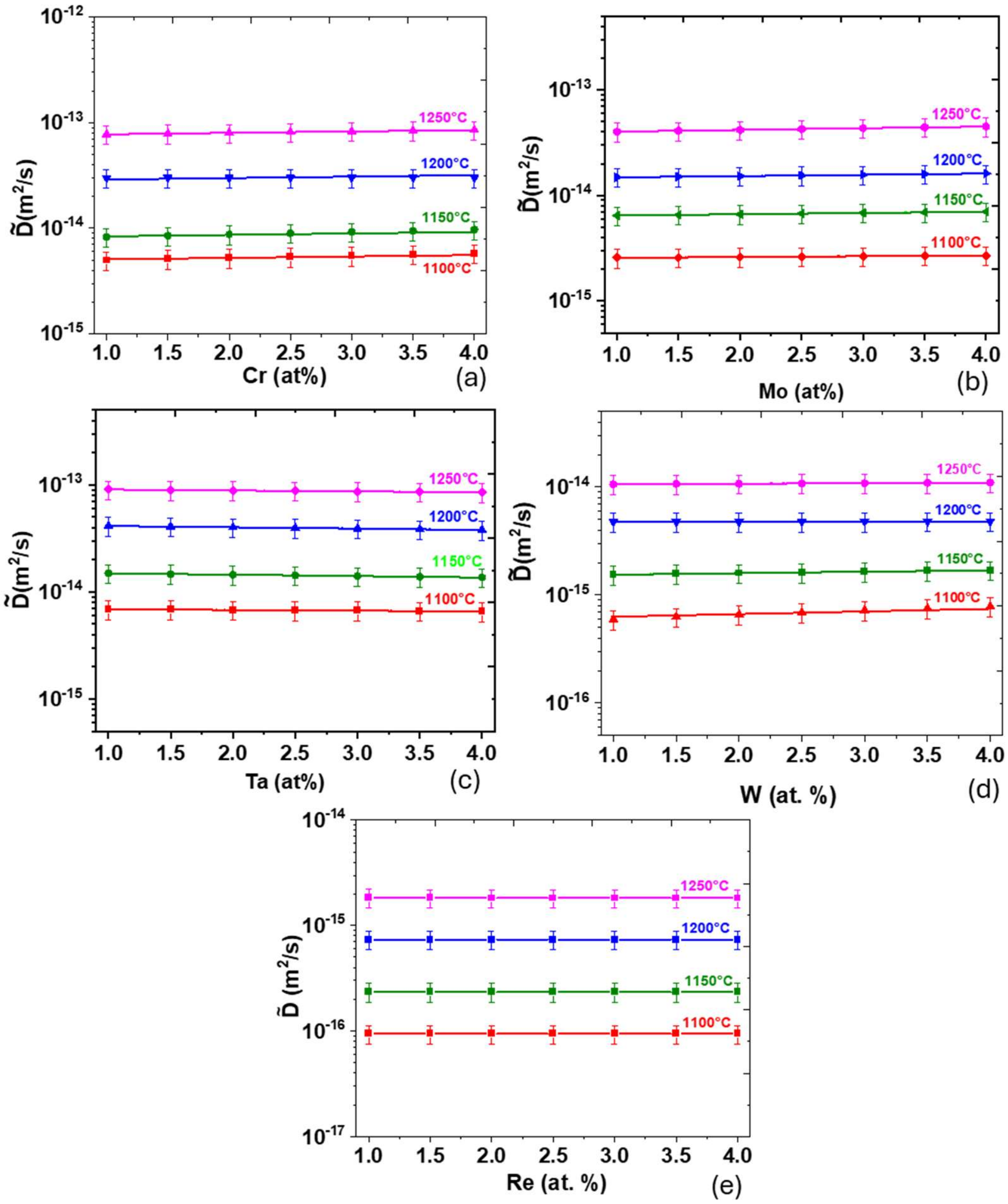

Fig. 2 Composition-dependent $\widetilde{D}$ in Ni-X (X = Cr, Mo, Ta, W, Re) in the temperature range of 1100 – 1250 ℃ (a) Ni-Cr, (b) Ni-Mo, (c) Ni-Ta, (d) Ni-W and (e) Ni-Re.

| X | 1100 °C | 1150 °C | 1200 °C | 1250 °C |
|---|---|---|---|---|
| | $\widetilde{D}$ ($m^2/s$) ($\times 10^{-15}$) | | | |
| Cr | 5.3 (4.7[1], 4.0[2], 3.0[3], 0.7[4], 5.0[5]) | 9.0 (10.5[1],10.0[2], 7.2[3], 1.6[4], 11.4[5]) | 30.0 (25.0[2], 12.6[3], 3.5[4], 24.8[5]) | 81.0 (70.0[2], 21.6[3], 7.1[4], 51.6[5] ) |
| Mo | 2.8 (1.5[6], 3.1[7]) | 6.8 (5.3[7]) | 17.0 (15.0[6], 12.0[7]) | 40.6 (25.5[7]) |
| Ta | 6.7 (4.0[8], 6.4[9]) | 14.0 (14.1[9]) | 35.0 (20.0[8], 29.7[9], 20.0[10]) | 59.2 (53.1[8], 59.2[9]) |
| W | 0.7 (0.5[12], 0.7[17]) | 1.6 (1.6[11], 1.4[12], 1.8[17] ) | 4.8 (3.2[8],3.2[10],3.4[12], 4.3[17]) | 11.0 (20.0[8], 10[10], 8.6[11], 7.7[12], 9.7[17]) |
| Re | 0.1 (0.1[8], 0.5[13], 0.1[14]) | 0.3 (0.9[13], 0.3[14], 0.3[15]) | 0.7 (0.4[8] , 0.5[10], 1.4[14], 1.2[16]) | 1.8 (1.9[10], 1.2[15]) |
| Al | (13.0[18], 11.9[19]) | (30.0[18], 27.2[19]) | (62.0[18], 58.8[19]) | (130.0[18], 120.7[19]) |

Table 2 Comparison of $\widetilde{D}$ at the 2.5 at.% of X (X = Cr, Mo, Ta, W, Re, and Al) in the Ni-X systems as estimated in this study and the data available in the literature [1-19].

In all Ni–X binary systems, the estimated interdiffusion coefficients ($\widetilde{D}$) exhibit a linear dependence on composition at a given temperature, consistent with trends reported in earlier studies across various alloy systems [1–17]. Conversely, the thermodynamic factor ($\Phi = \Phi_{\text{Ni}} = \Phi_{\text{X}}$), defined as $\Phi_{\text{Ni}} = \frac{\partial \ln a_{\text{Ni}}}{\partial \ln N_{\text{Ni}}} = \Phi_{\text{X}} = \frac{\partial l \quad a_{\text{X}}}{\partial \ln N_{\text{X}}}$, increases with the concentration of X in these binary systems [40], as shown in Fig. 3. This variation is relatively modest for Cr, Mo, and Re, based on activity data extracted from the TCNI9 database of ThermoCalc [48].

At relatively low concentrations of X ($N_X \ll N_{\text{Ni}}$), the relationship simplifies to $\widetilde{D}(X) = N_X D_{\text{Ni}} + N_{\text{Ni}} D_X \approx D_X \approx D_X^* \Phi$, where $D_X$ and $D_{\text{Ni}}$ are the intrinsic diffusion coefficients, and $D_X^*$ is the tracer diffusion coefficient of X. This indicates that $D_i^*$decreases with increasing composition, though only marginally for Cr, Mo, and Re. Such differences between $\widetilde{D}$ and $D_i^*$ have recently been demonstrated in the Ni–Co binary system, where $\widetilde{D}$ was determined using the diffusion couple method and $D_i^*$ via the radiotracer method [49].

In contrast, the thermodynamic factors increase more significantly in the Ni–W and Ni–Ta systems. Since $\widetilde{D}$ does not vary substantially in these cases (similar trends are reported in Refs. [8–12,17]), this suggests a stronger decreasing tendency in $D_i^*$ for W and Ta with increasing composition, compared to Cr, Mo, and Re.

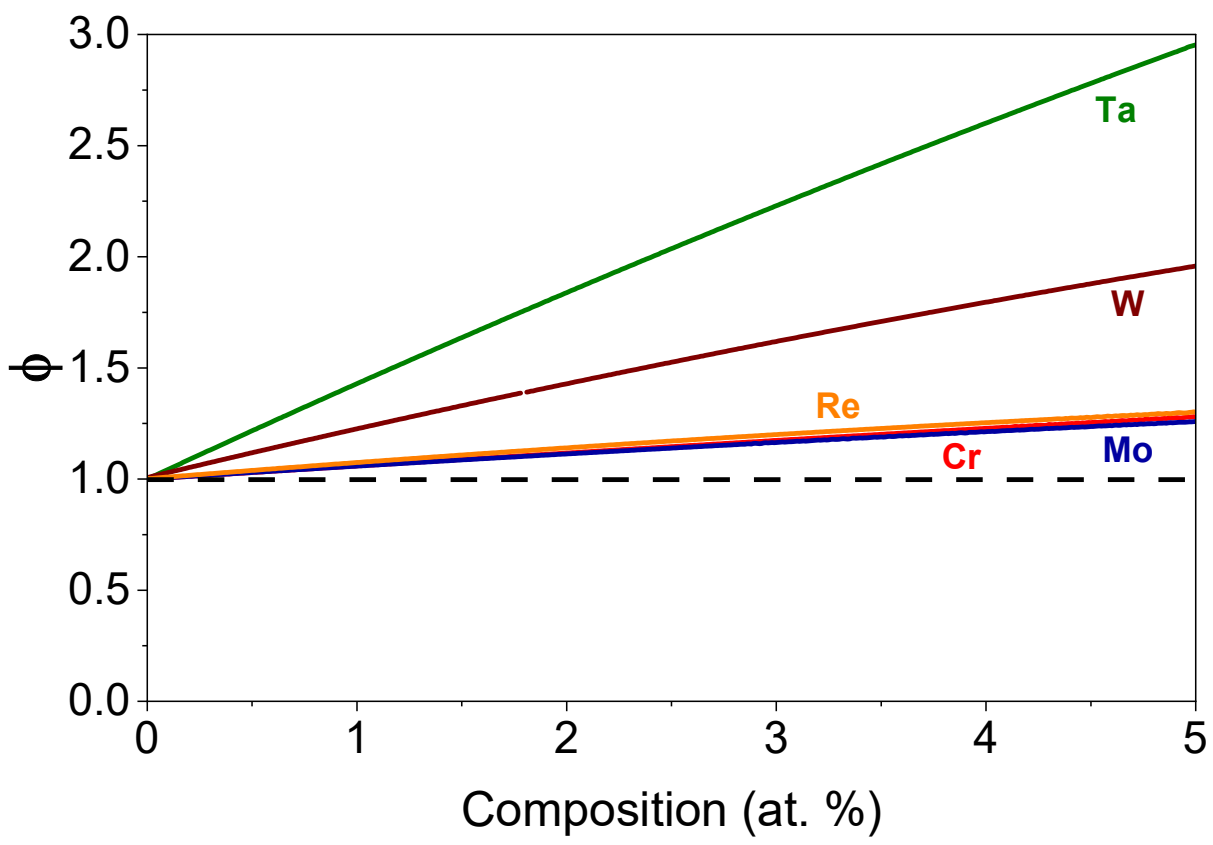

Fig. 3. Thermodynamic factors of elements X (= Cr, Ta, Mo, W, Re) at 1200 ºC.

The temperature-dependent values of $\widetilde{D}$at 2.5 at.% X in the Ni–X systems are presented in Fig. 4, together with literature data for the Ni–Al system [18,19]. The diffusion coefficients follow the trend $\widetilde{D}(\text{Re}) < \widetilde{D}(\text{W}) < \widetilde{D}(\text{Mo}) < \widetilde{D}(\text{Cr}) \approx \widetilde{D}(\text{Ta}) < \widetilde{D}(\text{Al})$. Although absolute values vary with composition and ordering of the systems, a comparable hierarchy of relative diffusion coefficients for Ni, Cr, and Al has also been reported in Ni–Co–Cr, Ni–Co–Fe–Cr, and Ni–Co–Fe–Cr–Al medium- and high-entropy alloys [50-56]. Among the investigated elements, W exhibits relatively smaller diffusion coefficients, while Re consistently shows the lowest values.

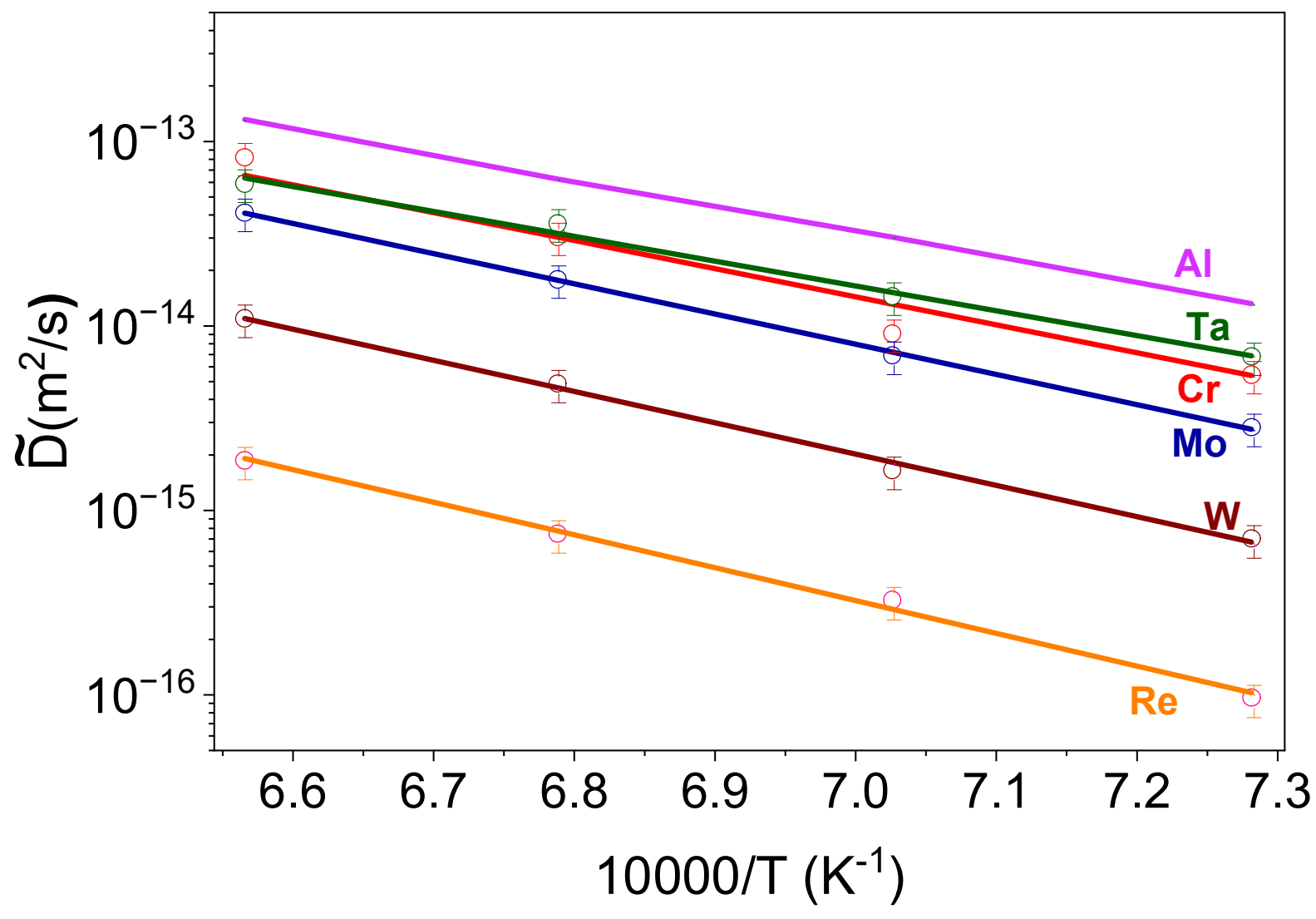

Fig. 4 The Arrhenius plot of $\widetilde{D}$ estimated at 2.5 at.%X in the Ni-X (Cr, Ta, Mo, W, Re) systems in the temperature range of 1100 - 1200 ºC. For reference, the $\widetilde{D}$ of the Ni-Al system at 2.5 at.% Al are reproduced from Ref. [18].

The activation energies were determined using the Arrhenius relation, $\widetilde{D} = \widetilde{D}_0 \exp(-Q/RT)$, where $\widetilde{D}_0$ is the pre-exponential factor, $Q$the activation energy, $R$the gas constant, and $T$the absolute temperature. A comparison of the calculated values for the Ni–X systems is presented in Fig. 5. Among the investigated elements, Re exhibits the highest activation energy, while Ta shows the lowest. Although the interdiffusion coefficients of Ta and Cr are similar (see Figure 4), their activation energies differ by nearly 30 kJ/mol. The relative ordering of activation energies obtained in this study follows $Q_{\mathrm{Re}} > Q_{\mathrm{W}} > Q_{\mathrm{Mo}} > Q_{\mathrm{Cr}} > Q_{\mathrm{Ta}}$. These results are broadly consistent with values reported in the literature (see supplementary file for comparison). For the Ni–Re system in particular, multiple studies have reported activation energies spanning a wide range from 196 to 412 kJ/mol, with the closest agreement found with the data of Chen et al. [10].

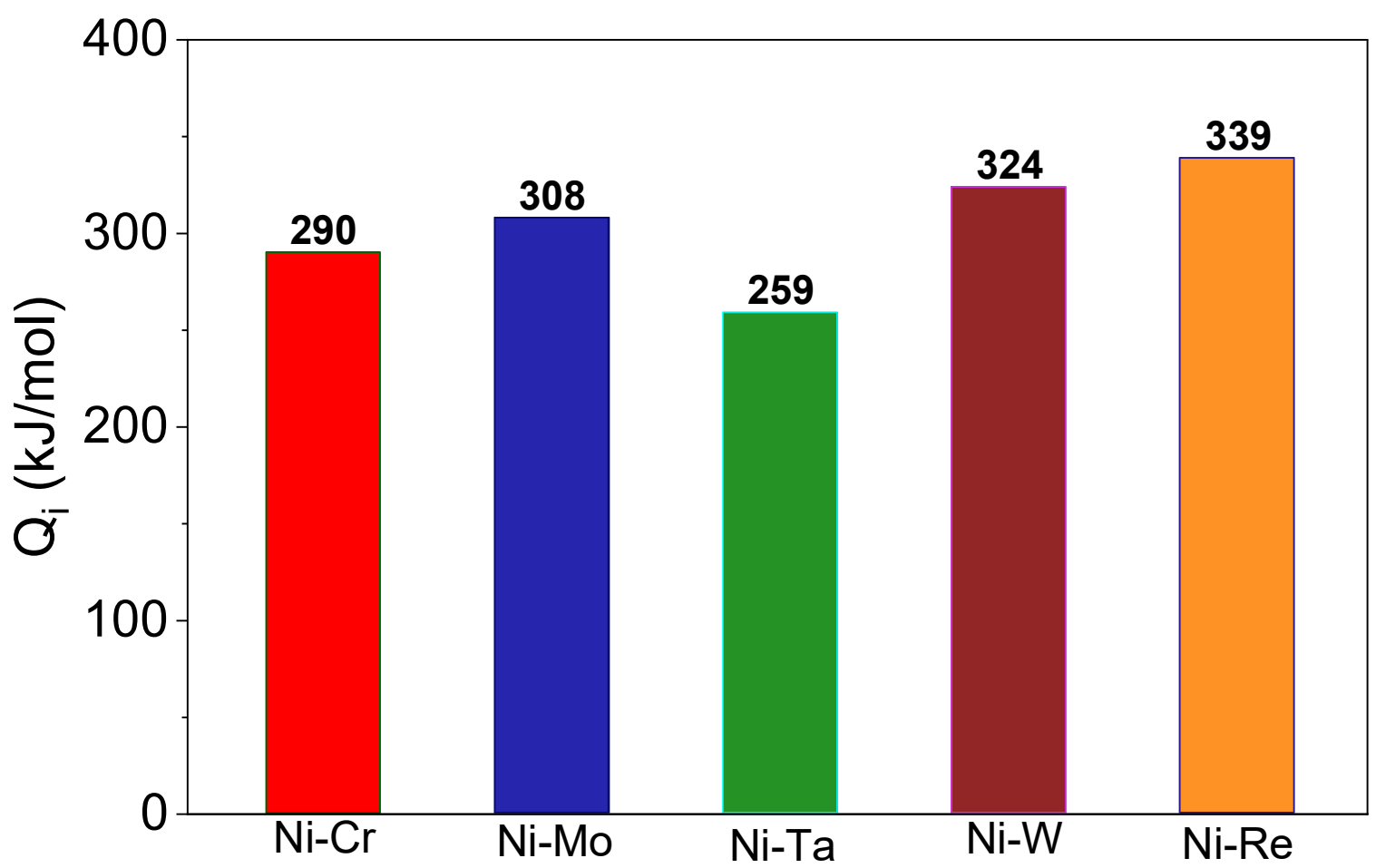

Fig. 5 Activation energies for interdiffusion calculated (with roughly ±5 % error across the systems) in the Ni-X systems.

*3.1.2 First-principles calculations of the activation energies for impurity diffusion coefficients of elements X (= Cr, Ta, Mo, W, Re) in Ni*

First-principles calculations were employed to estimate relative activation energies and to validate the experimentally determined values. However, consistent with earlier investigations in similar alloy systems, the calculated values do not always align precisely with experimental data, a discrepancy also noted in previous studies [57-64]. Furthermore, determining activation energies for diffusing species in binary or higher-order alloys using this approach is inherently complex.

Since the variation of $\widetilde{D}$ in binary systems is linear and composition-independent (within the composition range considered in this study); the impurity diffusion coefficient of X in Ni, $D_{X(Ni)}^{imp}$ (X = Cr, Mo, Ta, W, Re) in Ni can be approximated by extrapolating $\widetilde{D}$ to zero solute concentration. This allows the activation energy for impurity diffusion to be evaluated within this framework. Fig. 6a schematically illustrates the atomic configuration used to calculate the activation energy for the substitutional diffusion of impurity element X. In this approach, the activation energy is expressed as $Q = E_V^X + E_m^X$, where $E_V$ is the vacancy formation energy and $E_M^X$ the migration energy of the impurity element [59,60,62].

The vacancy formation energies in the presence of solute elements X, calculated using Eq. 1, are shown in Fig. 6b. These values are similar across different systems and closely match the formation energy of pure Ni (163 kJ/mol). Consequently, variations in $E_m^X$ emerge as the primary factor governing the differences in $Q$ observed among the Ni–X binaries.

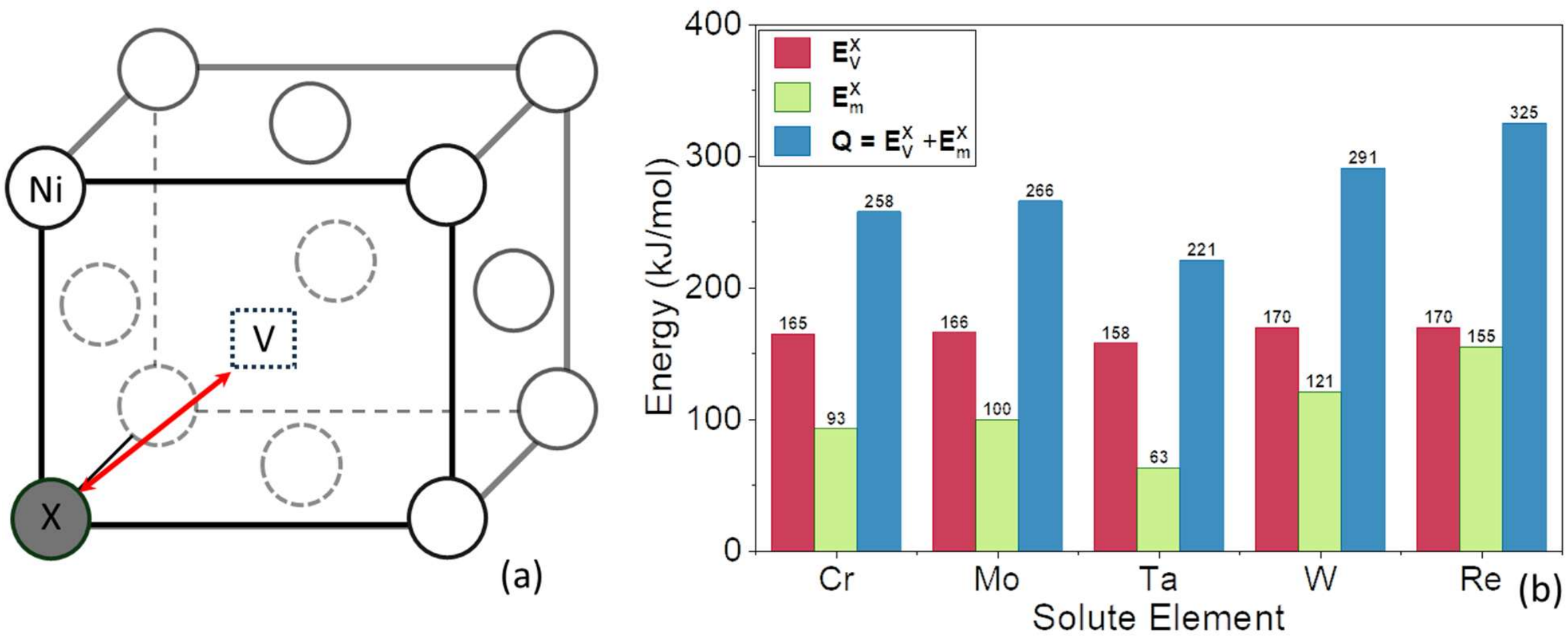

Fig. 6 (a) A schematic representation of FCC Ni showing the presence of element X (Cr, Mo, Ta, W, Re) next to a vacancy. (b) The calculated energy for vacancy formation $E_V^X$, energy for the migration of X to the position of vacancy ($E_M^X$) and the activation energy for diffusion, $Q_X (= E_V^X + E_M^X)$.

*3.2 Diffusion analysis in the ternary Ni-Al-X systems*

*3.2.1 Experimental estimation of the diffusion coefficients*

As previously discussed, the interdiffusion coefficient ($\widetilde{D}_{ij}^n$) can be directly evaluated at the intersecting composition using Eq. 11. The relationship between interdiffusion and intrinsic diffusion coefficients is expressed as

$$\widetilde{D}_{ij}^{n} = D_{ij}^{n} - N_i \sum_{k=1}^{n} D_{kj}^{n} \tag{19}$$

The intrinsic diffusion coefficients ($D_{ij}^{n}$) are, in turn, related to the tracer diffusion coefficients ($D_i^*$) of the elements—neglecting the vacancy wind effect—by excluding the second term in Eq. 6 [65–67]:

$$D_{ij}^{n} = \frac{N_i}{N_j} D_i^* \, \Phi_{ij}^{n} \tag{20}$$

Here, $\Phi_{ij}^{n}$ denotes the thermodynamic factor, defined in terms of activities as

$$\Phi_{ij}^{n} = \left(\frac{\partial \mathrm{l} \quad a_i}{\partial \ln \ N_1}\right) - \frac{N_1}{N_n}\left(\frac{\partial \ln \ a_i}{\partial \ln \ N_n}\right). \tag{21}$$

The vacancy wind contribution is neglected, as its influence is marginal at the relatively low concentrations of X and Al considered in these calculations.

As proposed by Kirkaldy and Lane [68], once the four interdiffusion coefficients ($\widetilde{D}_{ij}^{n}$) are estimated, they can be correlated with Eqs. 19 and 20 to establish relationships with the tracer diffusion coefficients ($D_i^*$). Using a least-squares approach, three tracer diffusion coefficients can thereby be determined. The intrinsic diffusion coefficients ($D_{ij}^{n}$) are then obtained from Eq. 20, providing the complete set of diffusion parameters required to elucidate different aspects of the diffusion process.

An alternative procedure, applicable even to higher-order systems [50–52], involves substituting Eqs. 19 and 20 directly into Eq. 11 with expressions that relate the interdiffusion fluxes to $D_i^*$. This method enables direct evaluation of $D_i^*$ from interdiffusion fluxes via the least-squares method, followed by calculation of $D_{ij}^{n}$ using Eq. 20 and $\widetilde{D}_{ij}^{n}$ using Eq. 19. This latter approach has been adopted in our analysis and calculations.

The intersecting diffusion profiles at 1200 °C for all Ni–Al–X systems (X = Cr, Mo, Ta, W, Re) are presented in Fig. 7. Overall, the diffusion profiles exhibit similar characteristics across the different systems, with variations primarily reflected in the diffusion lengths, which depend on the diffusion rates of the constituent elements. An example of the profiles obtained in the Ni–Al–Mo system is shown in Fig. 8.

The intersecting compositions, identified from the Gibbs triangle plots (Fig. 7), differ among systems due to variations in relative diffusivities. Within a given system, however, only minor or negligible differences are observed across temperatures. In all systems except Ni–Al–Re, the diffusion paths intersect well inside the interdiffusion zone, away from the end members. This is advantageous for calculating composition gradients, which are essential for reliable estimation of diffusion coefficients.

In contrast, the Ni–Al–Re system exhibits intersections very close to the end-member compositions, owing to the pronounced disparity in diffusivities between Al and Re. Similar behavior has been reported previously [25,27]. Although diffusion coefficients were estimated in those studies despite these limitations, we identified reliability concerns arising from uncertainties in the determination of the composition gradient. Consequently, the intersecting diffusion couple method has been applied to all Ni–Al–X systems (X = Cr, Mo, Ta, W), but not to Ni–Al–Re. For the latter, diffusion coefficients were instead determined using a recently proposed method of direct estimation at the Kirkendall marker plane from a single diffusion profile. The position of this marker plane in one of the diffusion couples is indicated in Fig. 7 and discussed in detail later.

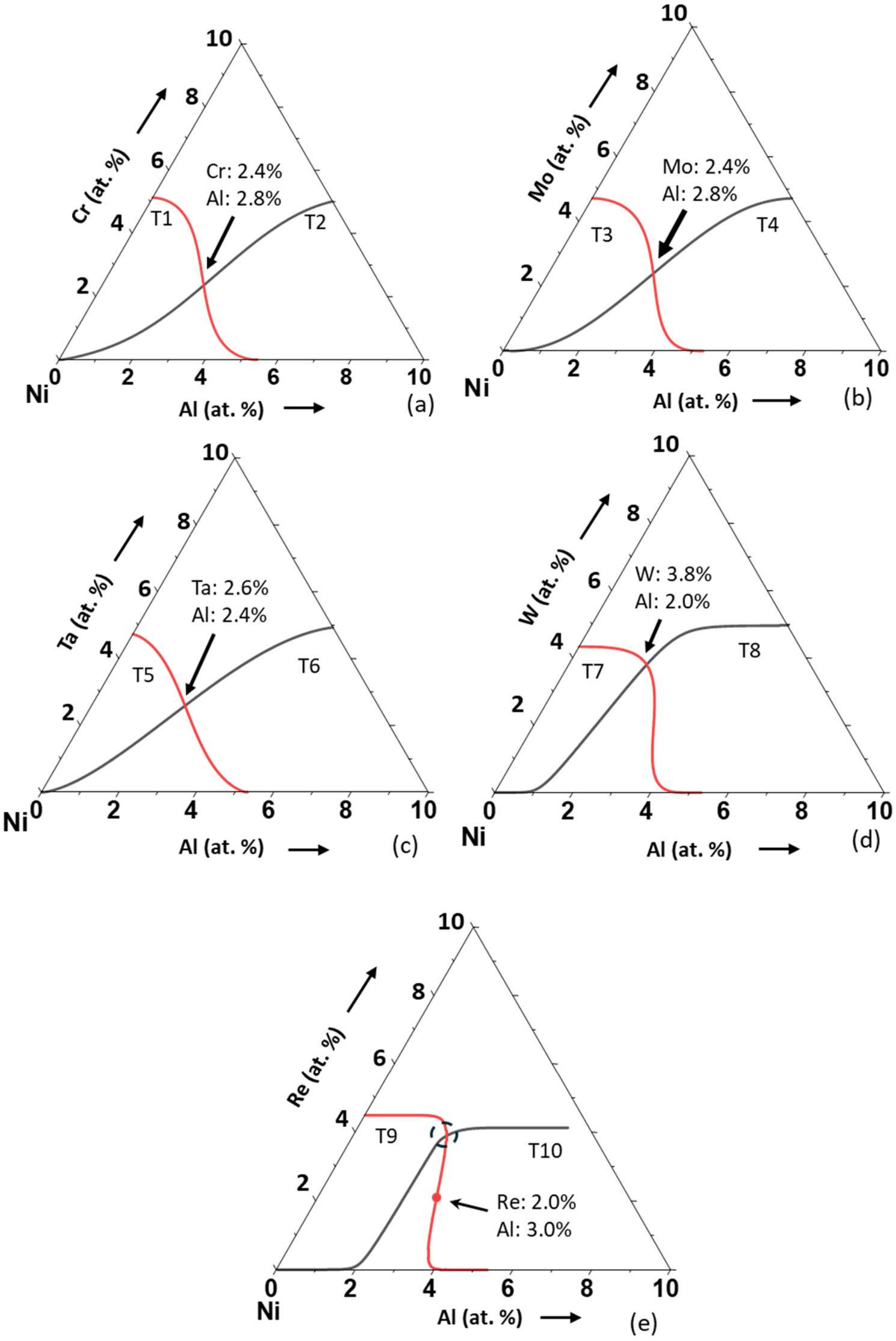

Fig. 7 Intersecting diffusion paths on Gibbs triangle in (a) Ni-Al-Cr, (b) Ni-Al-Mo, (c) Ni-Al-Ta, (d) Ni-Al-W and (e) Ni-Al-Re systems at 1200 °C. For the Ni-Al-Re system, we employed a method that directly calculates $D_i^*$ at the Kirkendall marker plane (the composition is shown as a filled circle).

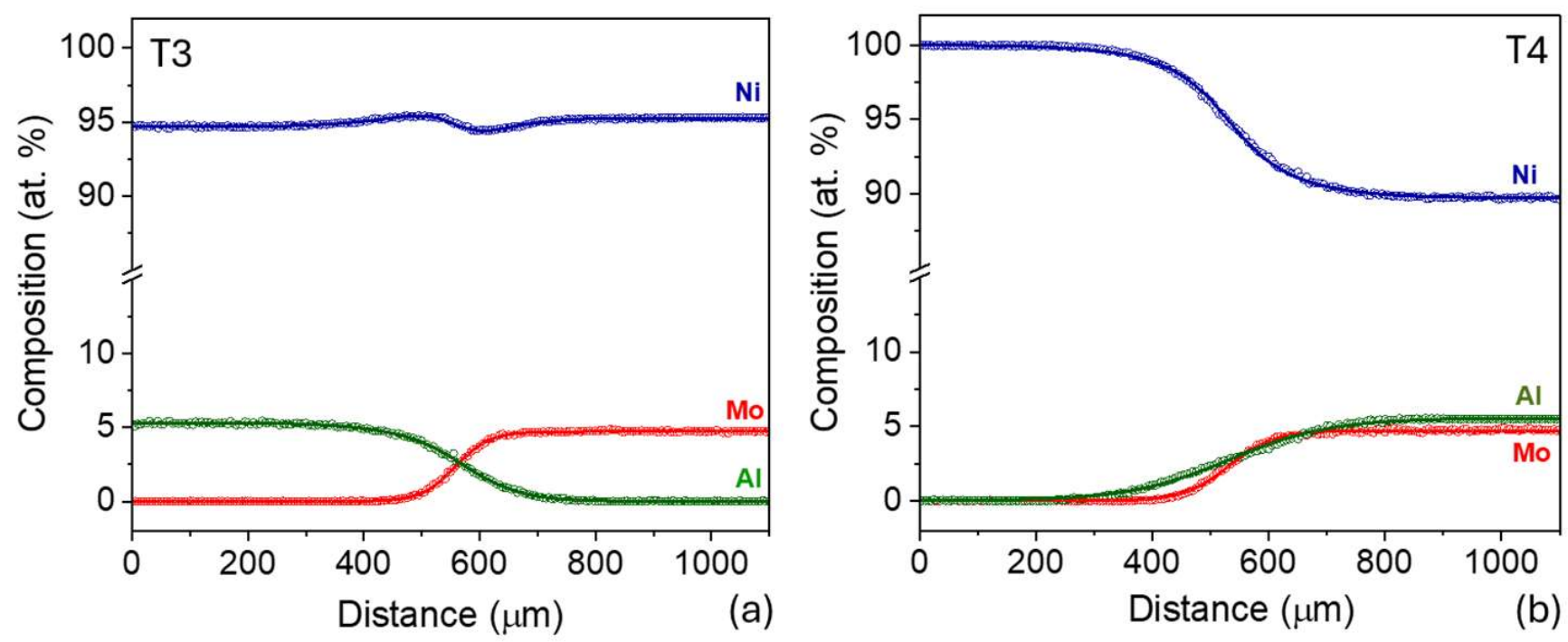

Fig. 8 The diffusion profiles produced at 1200 °C by annealing for 50 h in the Ni-Al-Mo system by coupling (a) Ni5Al and Ni5Mo (T3) and (b) Ni and Ni5Al5Mo (T4).

In the present analysis, tracer diffusion coefficients ($D_i^*$) were determined only for X and Al, as the interdiffusion process within this compositional regime is predominantly governed by the diffusion of minor constituents. This follows directly from Eq. 19, which, when expressed for a ternary system, reduces to

$$\widetilde{D}_{ij}^n = (1 - N_i)D_{ij}^n - N_i(D_{jj}^n + D_{nj}^n) \approx D_{ij}^n \tag{22}$$

under the condition that $N_i, N_j \ll N_n$, where $N_i$ and $N_j$ correspond to X and Al, and $N_n$ to Ni. Attempts to evaluate $D_{Ni}^*$, under these circumstances, frequently yield non-physical values, including unrealistically high, low, or even negative coefficients [29,69]. Accordingly, the self-diffusion coefficient of Ni (i.e., the tracer diffusion coefficient of Ni in pure Ni) was employed in the present calculations [70]. This substitution does not include a notable error in the estimated values of X and Al. In contrast, intermetallic compounds may exhibit pronounced sensitivity to minor compositional variations, owing to significant differences in point-defect concentrations that mediate substitutional diffusion [71–74].

The values of $D_{\mathrm{Ni}}^*$ in pure Ni adopted for the present study are: $2.9 \times 10^{-15}$ m$^2$/s at 1100 °C, $6.9 \times 10^{-15}$ m$^2$/s at 1150 °C, $15.3 \times 10^{-15}$ m$^2$/s at 1200 °C, and $32.4 \times 10^{-15}$ m$^2$/s at 1250 °C. The calculated tracer diffusion coefficients of Al and X, together with the intrinsic ($D_{ij}^n$) and interdiffusion ($\widetilde{D}_{ij}^n$) coefficients, are summarized in Table 3. Thermodynamic factors derived from activity data obtained using the ThermoCalc database TCNI9 [48,75] are given in the supplementary file.

Table 3 The $D_i^*$, $D_{ij}^n$ and $\widetilde{D}_{ij}^n$ calculated in the temperature range of 1100 – 1200 ºC in the Ni-Al-X (X= Cr, Mo, Ta and W) and 1150 – 1250 ºC in the Ni-Al-Re system. $D_{Ni}^* = 2.9 \times 10^{-15}$ (1100 ºC), $6.9 \times 10^{-15}$ (1150 ºC) and $15.3 \times 10^{-15}$ (1200 ºC) $32.4 \times 10^{-1}$ $m^2/s$ (1250 ºC) [70].

| Temperature (ºC) | Composition (at.%) | $D_i^*(\times 10^{-15}\ m^2/s)$ | | $D_{ij}^n(\times 10^{-15}\ m^2/s)$ | | | | | | $\widetilde{D}_{ij}^n(\times 10^{-15}\ m^2/s)$ | | | |
|---|---|---|---|---|---|---|---|---|---|---|---|---|---|
| | | $D_X^*$ | $D_{Al}^*$ | $D_{NiX}^{Ni}$ | $D_{NiAl}^{Ni}$ | $D_{XX}^{Ni}$ | $D_{XAl}^{Ni}$ | $D_{AlX}^{Ni}$ | $D_{AlAl}^{Ni}$ | $\widetilde{D}_{XX}^{Ni}$ | $\widetilde{D}_{XAl}^{Ni}$ | $\widetilde{D}_{AlX}^{Ni}$ | $\widetilde{D}_{AlAl}^{Ni}$ |
| 1100 | $NiCr_{2.5}Al_{2.6}$ | 3.9 | 9.6 | -4.0 | -3.9 | 4.5 | 0.8 | 2.1 | 10.8 | 4.5 | 0.6 | 1.4 | 10.6 |
| | $NiMo_{2.4}Al_{2.6}$ | 1.7 | 7.0 | -3.9 | -3.8 | 1.9 | 0.3 | 1.4 | 7.9 | 1.9 | 0.2 | 1.4 | 7.8 |
| | $NiTa_{2.1}Al_{2.4}$ | 4.6 | 14.0 | -6.3 | -4.2 | 8.1 | 1.6 | 5.6 | 15.4 | 8.0 | 1.4 | 5.4 | 15.1 |
| | $NiW_{4.4}Al_{1.6}$ | 0.4 | 6.3 | -6.3 | -5.0 | 0.7 | 0.2 | 1.5 | 6.7 | 0.9 | 0.2 | 1.6 | 6.7 |
| 1150 | $NiCr_{2.3}Al_{2.9}$ | 6.7 | 19.8 | -9.5 | -9.1 | 7.7 | 1.3 | 4.5 | 22.5 | 7.6 | 1.0 | 4.4 | 22.1 |
| | $NiMo_{2.4}Al_{2.9}$ | 4.3 | 18.8 | -9.3 | -9.1 | 4.9 | 1.1 | 4.1 | 21.4 | 4.9 | 0.8 | 4.1 | 21.0 |
| | $NiTa_{2.5}Al_{2.6}$ | 8.9 | 30.6 | -15.9 | -10.4 | 16.6 | 3.7 | 12.3 | 33.9 | 16.3 | 3.0 | 12.0 | 33.2 |
| | $NiW_{3.3}Al_{2.5}$ | 0.8 | 15.7 | -14.0 | -10.5 | 1.4 | 0.4 | 5.4 | 17.3 | 1.6 | 0.2 | 5.6 | 17.1 |
| | $NiRe_{2.0}Al_{3.0}$ | 0.1 | 17.0 | -9.8 | -10.2 | 0.2 | 0.03 | 4.7 | 20.5 | 0.3 | -0.2 | 4.8 | 20.2 |
| 1200 | $NiCr_{2.4}Al_{2.8}$ | 17.0 | 56.0 | -19 | -20 | 20 | 3 | 12 | 70 | 19 | 2 | 12 | 69 |
| | $NiMo_{2.4}Al_{2.8}$ | 8.0 | 35.0 | -20 | -20 | 9 | 1 | 7 | 39 | 9 | 1 | 7 | 39 |
| | $NiTa_{2.6}Al_{2.4}$ | 19.0 | 65.0 | -34 | -23 | 36 | 8 | 24 | 71 | 35 | 7 | 23 | 69 |
| | $NiW_{3.8}Al_{2.0}$ | 2.0 | 31.0 | -31 | -26 | 3 | 1 | 8 | 34 | 4 | 1 | 8 | 34 |
| | $NiRe_{2.0}Al_{3.0}$ | 0.3 | 33.0 | -21 | -23 | 0.3 | 0.06 | 9 | 40 | 0.5 | -0.3 | 9 | 39 |
| 1250 | $NiRe_{1.7}Al_{3.1}$ | 0.7 | 70.0 | -44.7 | -42.2 | 0.7 | 0.07 | 20.5 | 83.3 | 1.4 | -0.4 | 21.2 | 83 |

The diffusion coefficients in the Ni-Al-Re system are calculated at the Kirkendall marker plane using a recently established single-profile method. At the marker plane, the intrinsic fluxes of three elements $i$, $j$ and $n$ can be calculated to relate to the $D_{ji}^n$ by [40,43]

$$V_m J_i = -D_{ii}^n \frac{\partial N_i}{\partial x} - D_{ij}^n \frac{\partial N_j}{\partial x} \quad (23a)$$

$$V_m J_j = -D_{ji}^n \frac{\partial N_i}{\partial x} - D_{jj}^n \frac{\partial N_j}{\partial x} \quad (23b)$$

$$V_m J_n = -D_{ni}^n \frac{\partial N_i}{\partial x} - D_{nj}^n \frac{\partial N_j}{\partial x} \quad (23c)$$

The intrinsic flux can be calculated directly from the composition profile at the marker plane from [47,76,77]

$$V_m J_i = -\frac{1}{2t}\left[N_i^+ \int_{x^{-\infty}}^{x_K} Y_i dx - N_i^- \int_{x_K}^{x^{+\infty}} (1 - Y_i)\, dx\right] \quad (24)$$

By substituting Eq. 20 into Eq. 23, the intrinsic fluxes can be expressed directly in terms of the tracer diffusion coefficients ($D_i^*$). Consequently, $D_i^*$ of three elements can be determined from the three independent equations, from which the intrinsic diffusion coefficients ($D_{ij}^n$) and the interdiffusion coefficients ($\widetilde{D}_{ij}^{n}$) are subsequently obtained. The diffusion profile employed for this calculation, together with the position of the Kirkendall marker plane, is presented in Fig. 9. This profile is shown at the highest temperature investigated, as the extended diffusion length of Re facilitates clearer visualization of the diffusion characteristics.

A notable advantage of this profile is that both Al and Re exhibit zero composition in one of the terminal alloys. Under these conditions, Eq. 24 simplifies to $V_m J_i = -\frac{1}{2t}\left[N_i^{+}\int_{x^{-\infty}}^{x_K} Y_i\ dx\right]$(for Al) or $V_m J_i = \frac{1}{2t}\left[N_i^{-}\int_{x_K}^{x^{+\infty}}(1-Y_i)\ dx\right]$(for Re), depending on the plotting in Fig. 9. In such representations, $\partial N_{\mathrm{Al}}/\partial x$ is positive and $\partial N_{\mathrm{Re}}/\partial x$ is negative, ensuring positive values of $D_i^*$. If the plotting orientation is reversed, the signs of the intrinsic fluxes and composition gradients also reverse, yielding consistent results.

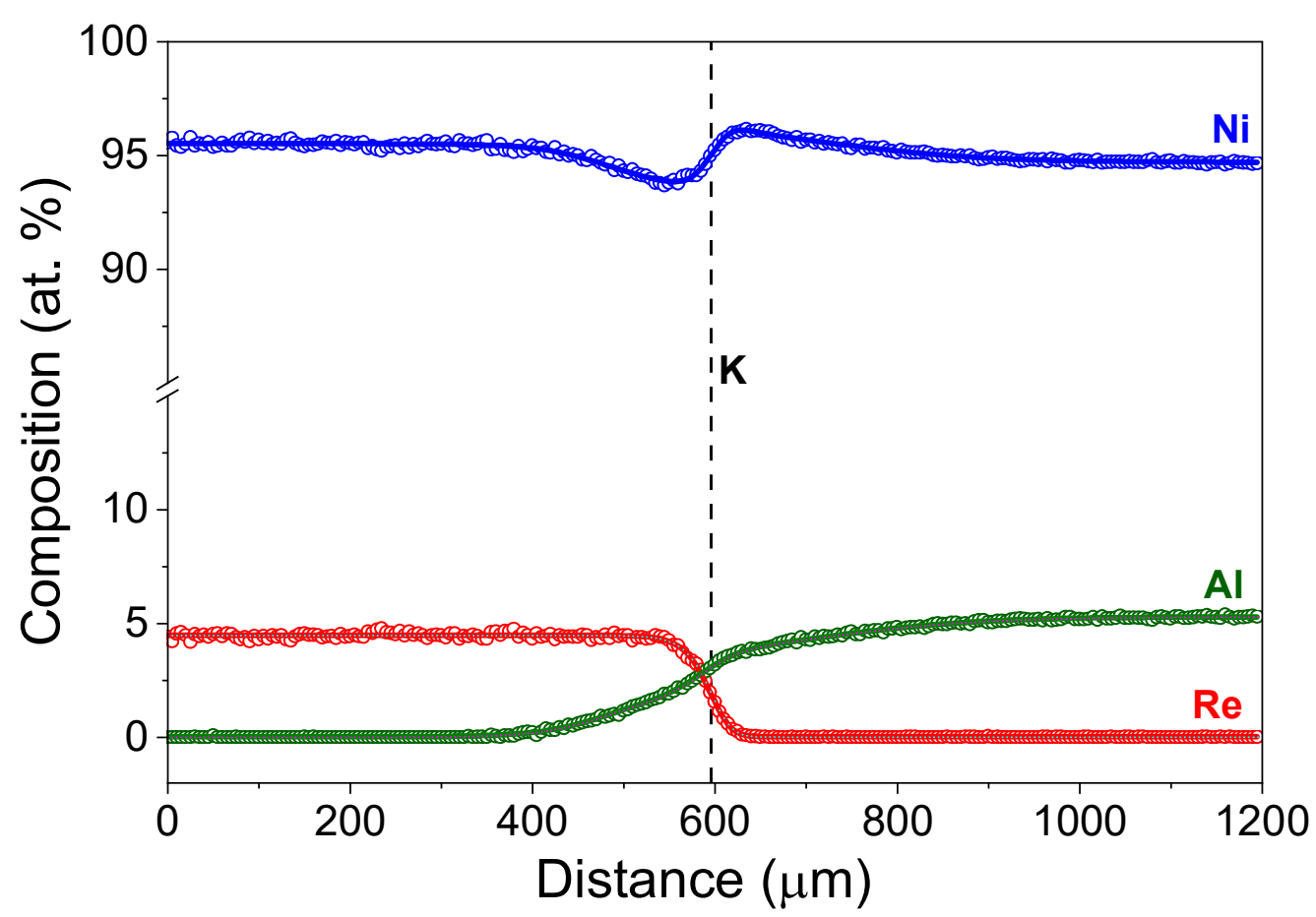

Fig. 9 Ni-Re-Al interdiffusion profile by coupling Ni-5Re and Ni-5Al alloys at 1250 °C annealed for 50h. The location of the Kirkendall marker plane (K) is shown.

Because of the zero composition at one end, the calculation errors associated with $D_i^*$ calculation is relatively small compared to those for Ni, where both $N_{\mathrm{Ni}}^{-}$ and $N_{\mathrm{Ni}}^{+}$ are non-zero. Extensive discussions of such errors are available in Refs. [29,69]. To mitigate this issue, only the $D_i^*$ of Al and Re are calculated, while the self-diffusion coefficient of Ni is employed, which introduces negligible error in the estimation of $\widetilde{D}_{ij}^{n}$ as the interdiffusion process in the Ni-rich region is dominated by diffusion of Al and Re.

A second diffusion profile, generated by coupling Ni with Ni5Al5Re (see the supplementary information), also benefits from zero Al and Re compositions at one end. However, the non-monotonic nature of Al diffusion in this profile, particularly the weak gradient near the Kirkendall plane, increases the likelihood of calculation errors. An alternative profile without these limitations was therefore preferred.

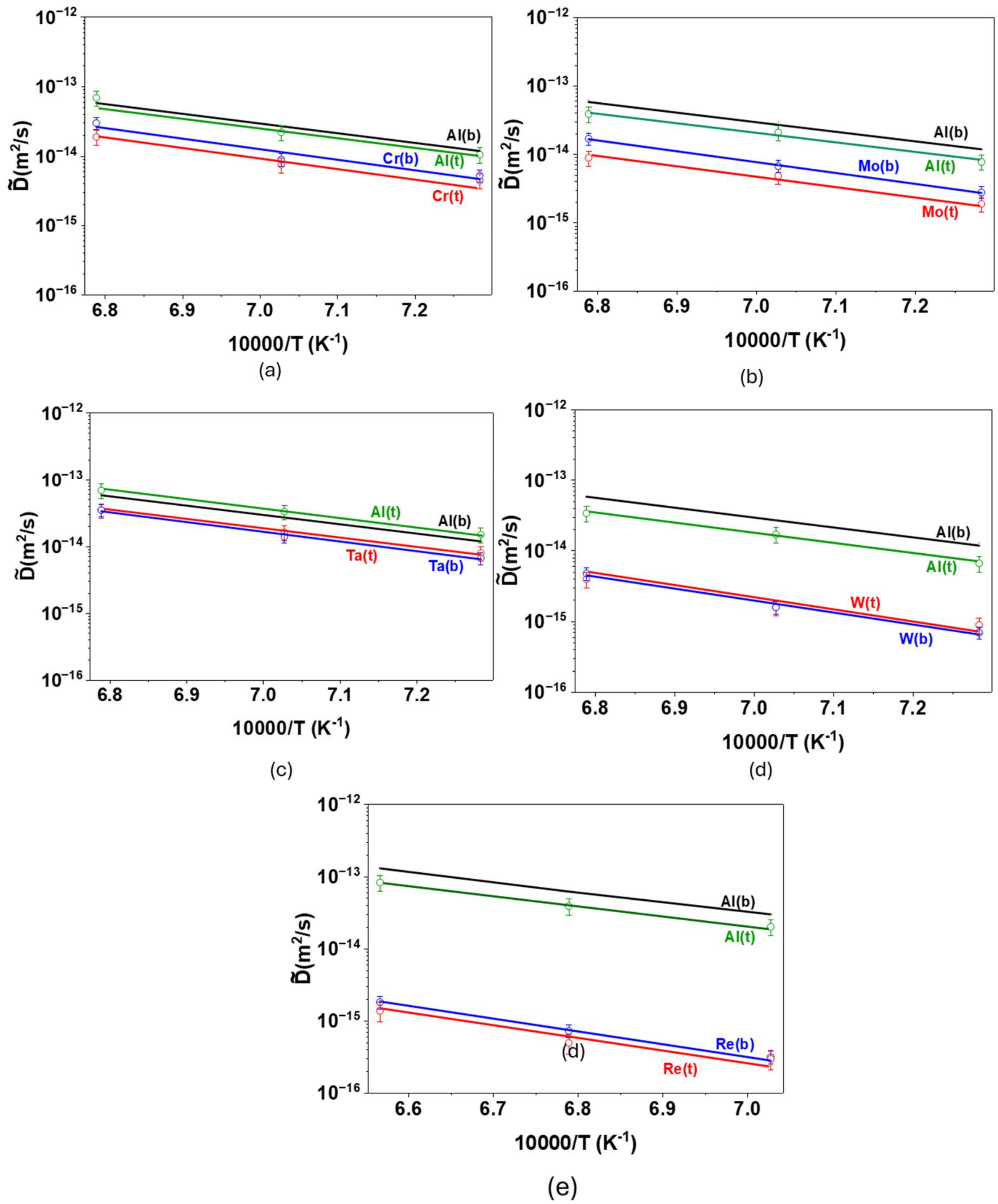

Fig. 10 The comparison of the main interdiffusion coefficients of X, $\widetilde{D}_{XX}^{Ni}$ (X= Cr, Mo, Ta, W, Re) and Al ($\widetilde{D}_{AlAl}^{Ni}$) in the ternary systems in comparison to the $\widetilde{D}$ in the Ni-X (X= Cr, Mo, Ta, W, Re, Al) systems: (a) Ni-Al, Ni-Cr and Ni-Cr-Al, (b) Ni-Al, Ni-Mo and Ni-Mo-Al, (c) Ni-Al, Ni-Ta and Ni-Ta-Al, (d) Ni-Al, Ni-W and Ni-W-Al and (e) Ni-Al, Ni-Re and Ni-Re-Al.

In this ternary Ni–Al–Re system, diffusion coefficients were evaluated over the temperature range 1150–1250 °C, which is 50 °C higher than the range used for other Ni–Al–X systems, owing to the significantly shorter diffusion length of Re compared to Al. For consistency, diffusion coefficients in all ternary systems were determined at three temperatures. The main interdiffusion coefficients were found to be comparable to those in the corresponding binary systems, with similar slopes in the Arrhenius plots, indicating comparable activation energies (Fig. 10) of 339 kJ/mol in contrast to the values reported by Mabruri et al. [27] (271 kJ/mol) and Chen et al. [26] (445 kJ/mol). Such deviations, either higher or lower, are unlikely compared with the binary data, reinforcing the reliability of the present analysis.

The difference between the main interdiffusion coefficient of Al between different Ni-Al-X systems can be understood from the plot, as shown in Fig. 11. The $\widetilde{D}$ calculated in the Ni-Al binary system is also shown for comparison. By Al(X), we mean the main interdiffusion coefficient of Al, $\widetilde{D}_{AlAl}^{Ni}$ in the Ni-Al-X ternary system and Al(b) indicates the $\widetilde{D}$ in the Ni-Al binary system. It can be seen that $\widetilde{D}_{AlAl}^{Ni}$ in ternary Ni-Al-Ta, Ni-Al-Cr, Ni-Al-Mo and Ni-Al-W follows the trend of the main interdiffusion coefficient of X, such that we have Al(Ta)>Al(Cr)>Al(Mo)>Al(W). However, the differences are not as high as the difference between the main interdiffusion coefficients of X in different ternary systems, which can be realized from the $\widetilde{D}$ calculated in the binary Ni-X system, as shown in Fig. 3, since the main interdiffusion coefficients of X ($\widetilde{D}_{XX}^{Ni}$) in ternary systems have similar values compared to the binary $\widetilde{D}$ of these elements in the binary system. In the Ni-Al-Ta system, $\widetilde{D}_{AlAl}^{Ni}$ is found to be slightly higher than the $\widetilde{D}$ in the binary Ni-Al system. On the other hand, $\widetilde{D}_{AlAl}^{Ni}$ is not the lowest in the Ni-Al-Re system (with similar values in Ni-Al-Mo, Ni-Al-W and Ni-Al-W systems), although $\widetilde{D}_{ReRe}^{Ni}$ is the lowest compared to $\widetilde{D}_{XX}^{Ni}$ in other systems. This is the reason that the differences between $\widetilde{D}_{AlAl}^{Ni}$ and $\widetilde{D}_{XX}^{Ni}$ increases in the order of Ni-Al-Ta, Ni-Al-Cr, Ni-Al-Mo, Ni-Al-W, Ni-Al-Re systems, which can be understood from the plots in Fig. 11. Therefore $\widetilde{D}_{AlAl}^{Ni}$ does not change in the same order compared to X in different Ni-Al-X systems.

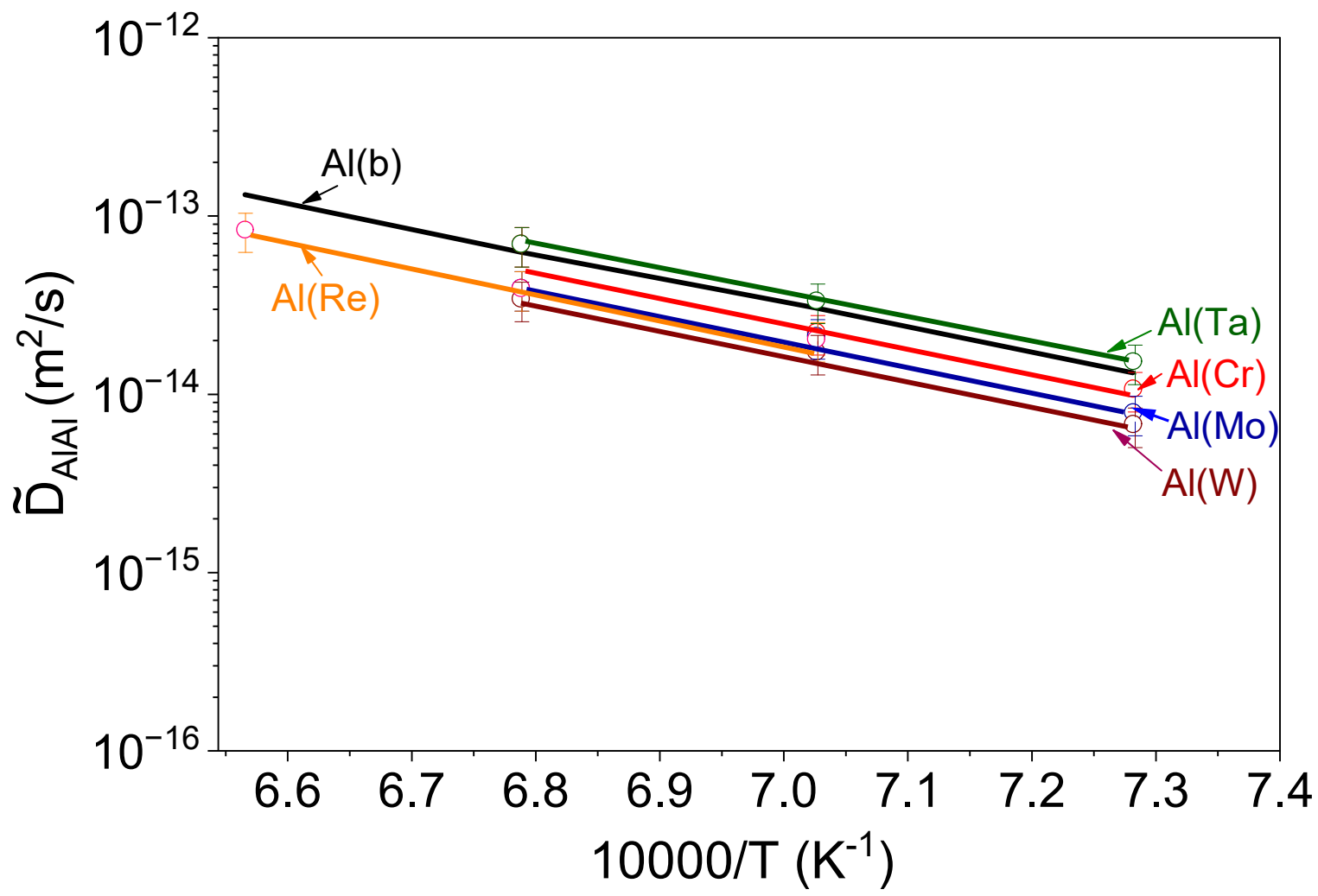

Fig. 11 The $\widetilde{D}_{AlAl}^{Ni}$in the ternary Ni-Al-X (X= Cr, Mo, Ta, W, Re) systems, denoted by Al (X), in comparison to the $\widetilde{D}$ in the binary Ni-Al system, denoted by Al(b).

The activation energies of X and Al in both binary and ternary systems are comparable, as indicated by their similar gradients within experimental error. Determining activation energies using ab initio methods is highly complex due to the vast number of possible atomic configurations. Experimentally, however, we measure the bulk activation energy, which averages across all configurations. Since the concentrations of Al and X are relatively small, the overall diffusion process closely resembles that of the binary Ni–X and Ni–Al systems. Nevertheless, in certain cases, the diffusion of X or Al can be influenced by the presence of another element.

To better understand this effect, without directly linking to experimentally derived values, we consider a specific configuration in which both X and Al are located adjacent to a vacancy (Fig. 12a). In this scenario, we calculated the migration energies for X as it exchanges positions with the vacancy. The results (Fig. 12b) show that the migration energy increases by approximately 15–30 kJ/mol. In contrast, the migration energy of Al is only marginally affected by X, with differences ≤ 5 kJ/mol. This variation falls within the numerical uncertainty of the calculations, which is comparable to the convergence criterion used for structural relaxation of the supercells (0.05 eV, or about 5 kJ/mol). Thus, the influence of X on Al diffusion is negligible within the limits of computational accuracy.

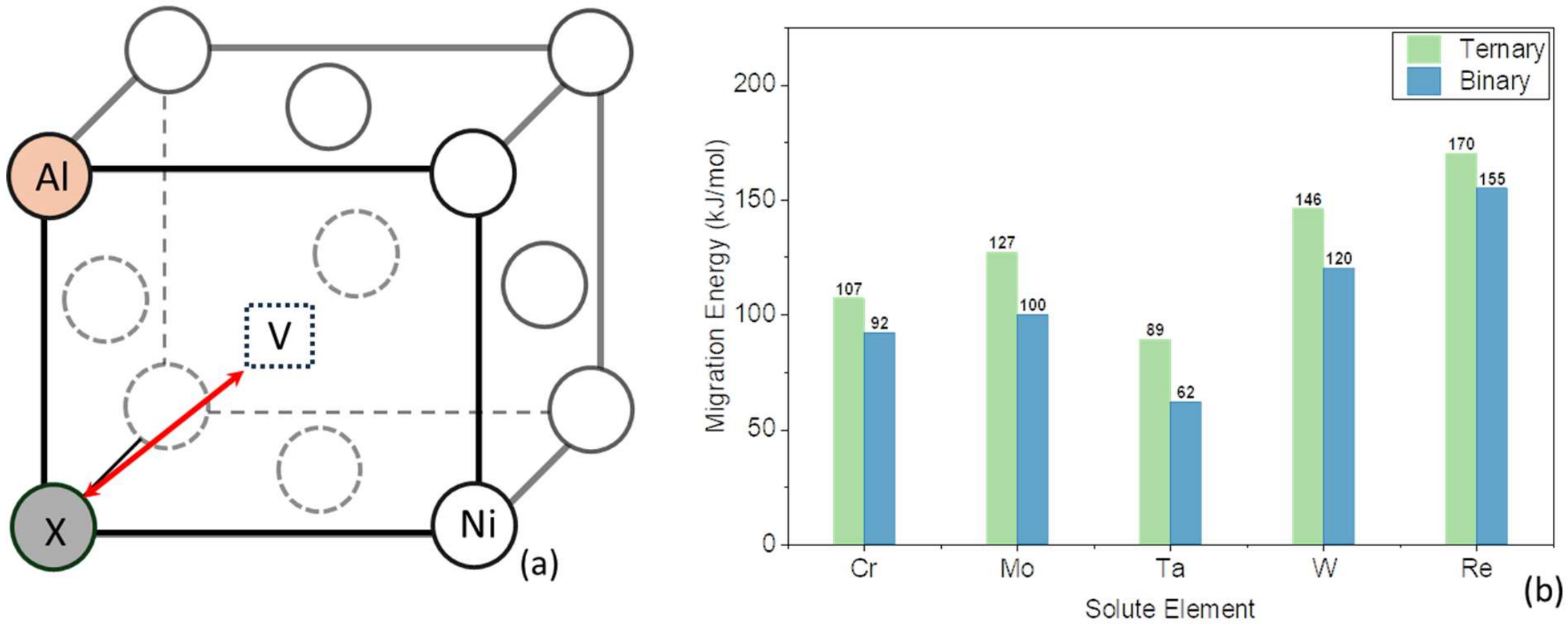

Fig. 12 (a) One of the possible atomic arrangements in Ni-Al-X alloy, and (b) the activation energy for migration of X in the presence of Al in comparison to binary Ni-X systems.

The available data within the temperature range of this study for selected systems are summarized in Table 4. Notable discrepancies are evident in the Ni–Cr–Al system [19], where inconsistencies arise not only in the absolute values but also in the relative magnitudes of the main and cross interdiffusion coefficients of Cr. Specifically, at 1100 °C, the reported cross-interdiffusion coefficient of Cr exceeded the main coefficient, whereas at 1200 °C, the opposite trend was observed, highlighting internal inconsistencies in the dataset. In contrast, our results show that the cross-interdiffusion coefficient of Cr remains consistently smaller than the main coefficients across all three investigated temperatures. Comparable magnitudes and signs of $\widetilde{D}_{ij}^{n}$ were also reported for the Ni–Al–Mo [21] and Ni–Al–Ta [23] systems, corroborating our findings. However, data reported for the Ni–Al–Ta system at 1100 °C in Ref. [22] differ substantially, particularly in the values and signs of $\widetilde{D}_{AlTa}^{Ni}$, from both our results and those of Ref. [23]. For the Ni–Al–Re system, diffusivities reported in Refs. [24,25,27] generally align with our data, though minor mismatches in certain coefficients remain. However, a comprehensive study spanning a temperature range, in comparison with data estimated in binary systems, was missing in all the systems reported to date, indicating differences in the diffusivities of X and Al across different Ni-Al-X systems.

Table 4: The $\widetilde{D}_{ij}^{n}$ reported in literature in the systems of our interest at certain temperatures [20-27]. Data beyond the temperature of interest are not included, for example, in the Ni-Al-W system [24,25].

| System | Composition | Temperature (ºC) | $\widetilde{D}_{ij}^{n}(\times 10^{-15}\ m^2/s)$ | | | |
|---|---|---|---|---|---|---|
| | | | $\widetilde{D}_{XX}^{Ni}$ | $\widetilde{D}_{XAl}^{Ni}$ | $\widetilde{D}_{AlX}^{Ni}$ | $\widetilde{D}_{AlAl}^{Ni}$ |
| Ni-Cr-Al | $NiCr_{6.74}Al_{3.21}$ [20] | 1100 | 8.9 | 10.2 | 3.5 | 16.1 |
| | $NiCr_{2.48}Al_{4.97}$ [20] | 1200 | 35.9 | 13.2 | 18.9 | 84.3 |
| Ni-Mo-Al | $NiMo_{2.56}Al_{2.48}$ [21] | 1100 | 1.3 | 1.2 | 1.4 | 1.6 |
| | $NiMo_{2.25}Al_{2.33}$ [21] | 1200 | 8.1 | 3.5 | 7.7 | 32.2 |
| Ni-Ta-Al | $NiTa_{1.71}Al_{1.11}$ [22] | 1200 | 15.7 | 1.2 | -0.3 | 30.6 |
| | $NiTa_{2.20}Al_{2.26}$ [23] | 1100 | 5.7 | 2.0 | 3.19 | 11.7 |
| | $NiTa_{2.17}Al_{2.28}$ [23] | 1200 | 33.3 | 11.2 | 14.9 | 76.5 |
| Ni-Re-Al | $NiRe_{1.25}Al_{3.50}$ [26] | 1200 | 0.3 | - | - | - |
| | $NiRe_{4.60}Al_{2.90}$ [27] | 1200 | 0.8 | -0.02 | 7.4 | 33 |
| | $NiRe_{4.60}Al_{2.50}$ [27] | 1250 | 2.0 | -0.8 | 13 | 64 |
| | $NiRe_{4.20}Al_{2.90}$ [24] | 1250 | 2.0 | -1.2 | 7.7 | 62 |
| | $NiRe_{4.30}Al_{2.50}$ [25] | 1250 | 2.2 | -1.2 | - | - |

The preceding discussion is based on the fact that $\widetilde{D}_{ij}^{n}$ represents the average of a set of three $D_{ij}^{n}$ in a ternary system, as expressed in Eq. 6. These coefficients are essential for establishing direct correlations with the diffusion profiles. Because they are estimated within relatively narrow compositional ranges of X and Al, the main interdiffusion coefficients are nearly equivalent to the corresponding intrinsic diffusion coefficients. In contrast, certain cross-interdiffusion coefficients, though smaller in magnitude, exhibit noticeable differences, while retaining consistent signs, with the exception of $\widetilde{D}_{ReAl}^{Ni}$ and $D_{ReAl}^{Ni}$. These are related by

$$\widetilde{D}_{ReAl}^{Ni} = (1 - N_{Re})D_{ReAl}^{Ni} - N_{Re}\left(D_{AlAl}^{Ni} + D_{NiAl}^{Ni}\right).$$

For instance, at 1250 °C, $D_{\mathrm{ReAl}}^{\mathrm{Ni}} = 0.07 \times 10^{-15}$, which is an order of magnitude smaller than $D_{ReRe}^{Ni} = 0.7 \times 10^{-15}$. This demonstrates that the influence of the cross-diffusion coefficient, despite its positive diffusional interaction, is negligible. By contrast, the calculated values of $\widetilde{D}_{\mathrm{ReRe}}^{\mathrm{Ni}} = 1 \times 10^{-15}$ and $\widetilde{D}_{\mathrm{ReAl}}^{\mathrm{Ni}} = -0.4 \times 10^{-15}$ indicate that the absolute magnitudes exceed those of the intrinsic coefficients, owing to contributions from other intrinsic terms ($D_{AlAl}^{Ni}$ and $D_{NiAl}^{Ni}$). Moreover, the negative value of $\widetilde{D}_{ReAl}^{Ni}$ suggests a strong negative diffusional interaction. Thus, reliance on the averaged value of $\widetilde{D}_{ReAl}^{Ni}$ misleadingly suggests the presence of such diffusional interaction, whereas the correct interaction is revealed only through the intrinsic coefficient $D_{\mathrm{ReAl}}^{Ni}$.

Such results, obtained through direct estimation of $\widetilde{D}_{ij}^{n}$, have also been reported in Refs. [24, 25, 27]. However, reliance solely on $\widetilde{D}_{ij}^{n}$ data, without reference to the intrinsic coefficients, can lead to misleading interpretations, such as an erroneously high negative interaction for Re (approximately 40% relative to the main interdiffusion coefficient). Historically, most studies following the diffusion couple method in ternary systems have reported only $\widetilde{D}_{ij}^{n}$, but the present analysis underscores the necessity of estimating and interpreting $D_{\mathrm{ij}}^{n}$, which more accurately reflect the diffusion characteristics of individual elements. Without such analysis, even the interactions of Ni with X and Al ($D_{\mathrm{NiX}}^{\mathrm{Ni}}, D_{\mathrm{NiAl}}^{\mathrm{Ni}}$) remain unclear, despite their significant role in defining the overall diffusion behavior, which is discussed next.

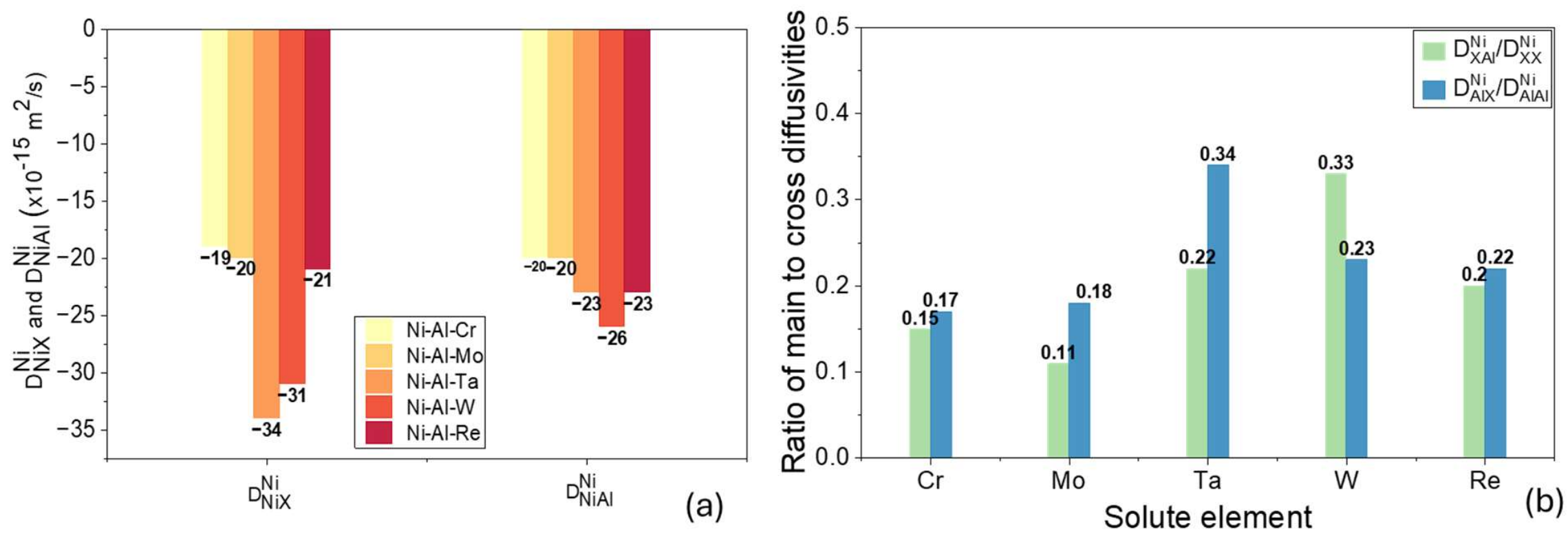

Fig. 13 (a) $D_{NiX}^{Ni}$ and $D_{NiAl}^{Ni}$ (b) the ratio of $D_{XAl}^{Ni}/D_{XX}^{Ni}$ and $D_{AlX}^{Ni}/D_{AlAl}^{Ni}$ in different Ni-Al-X systems at 1200 ºC.

At 1200 °C, the values of $D_{NiX}^{Ni}$ and $\mathrm{D}_{\mathrm{NiAl}}^{\mathrm{Ni}}$ are plotted in Fig. 13a. The results indicate that Ni exhibits its strongest diffusional interactions with both X (Ta) and Al in the Ni–Al–Ta system. While the values of $D_{NiAl}^{Ni}$ remain relatively consistent across different systems, significant variation is observed in $\mathrm{D}_{\mathrm{NiX}}^{\mathrm{Ni}}$. In particular, the negative diffusional interaction of Ni with Ta ($D_{NiTa}^{Ni}$) is the largest in the Ni–Al–Ta system, comparable to $D_{NiW}^{Ni}$ in the Ni–Al–W system. In other ternary systems, these values are found to be similar.

Fig. 13b presents the ratio of the cross coefficient $D_{XAl}^{Ni}$ to the main coefficient $D_{XX}^{Ni}$, which is highest in the Ni–Al–W system. This indicates a relatively stronger diffusional interaction of W with Al compared to other systems, whereas the lowest ratio is observed for Mo in the Ni–Al–Mo system. Conversely, the ratio of $D_{AlX}^{Ni}$ to $D_{AlAl}^{Ni}$ reveals that Al exhibits the strongest diffusional interactions in the Ni–Al–Ta system, and the weakest in the Ni–Al–Cr and Ni–Al–Mo systems.

The main interdiffusion coefficients ($\widetilde{D}_{XX}^{Ni}$) which closely approximate the intrinsic values ($D_{XX}^{Ni}$), provide insight into the similarities between interdiffusion coefficients reported for binary systems ($\widetilde{D}$). Importantly, cross-diffusion coefficients play a critical role in ternary systems, as they can either enhance or diminish fluxes depending on their sign and the relative direction of diffusion. For instance, both $D_{\mathrm{XAl}}^{\mathrm{Ni}}$ and $D_{\mathrm{AlX}}^{\mathrm{Ni}}$ are positive in all systems studied. This implies that fluxes are enhanced when X and Al diffuse in the same direction (see Fig. 8b) but reduced when they diffuse in opposite directions (see Fig. 8a).

Such contributions from cross-diffusion coefficients are additional to those from the main coefficients and are therefore essential for understanding flux behavior in ternary Ni–Al–X systems. These aspects cannot be captured solely through binary diffusion studies, which have dominated prior research in the Ni-based systems. The limited investigations on ternary systems and the absence of systematic comparisons across different alloy systems underscore the importance of extending diffusion studies to multicomponent systems.

The serpentine nature of diffusion paths observed on the Gibbs triangle (Fig. 7) can be rationalized by considering the relative diffusion rates of the constituent elements [78]. For example, in diffusion couples between Ni and Ni5Al5X (T2, T4, T6, T8, and T10 in Fig. 7), the paths deviate toward the X-rich side of the Ni5Al5X alloy because Al diffuses more rapidly than X. This preferential loss of Al enriches the alloy in X. To satisfy mass balance, the diffusion path must intersect the tie line connecting the end-member compositions, thereby bending back toward the Al-rich side near the Ni end member and producing a serpentine trajectory. A similar mechanism explains the serpentine paths in the Ni–Al–W (T7) and Ni–Al–Re (T9) systems, where diffusion couples between Ni5Al and Ni5X alloys show deviations toward higher Ni content on the Ni5Al side (due to rapid Al diffusion) and toward lower Ni content on the Ni5X side (due to faster Ni diffusion relative to X).

In contrast, the serpentine paths in Ni–Al–Cr (T1) and Ni–Al–Ta (T5) systems (Fig. 8) require more nuanced interpretation. Although Cr and Ta diffuse faster than Ni, the dominant effect arises from the Ni5Al end member, where Al diffuses more rapidly than Ni. This drives the path toward higher Ni content on the Ni5Al side, forcing the trajectory to bend toward lower Ni content on the Ni5Cr or Ni5Ta side to maintain mass balance. A comparable behavior is observed in the Ni–Al–Mo system (T3, Fig. 7), where Al diffuses significantly faster than both Ni and Mo. The uphill diffusion profile of Ni in the Ni5Al–Ni5Mo couple (Fig. 8a) illustrates this effect: the Ni profile bends toward higher Ni content on the Ni5Al side and toward lower Ni content on the Ni5Mo side.

The depth of serpentine deviation from the tie line is also governed by relative diffusion rates. In the Ni–Al–Re system (T9 and T10, Fig. 7), the extremely low diffusivity and short diffusion length of Re cause the paths near the end members to run nearly parallel to the Ni–Al axis, with only a narrow serpentine deviation in the central region where Re diffusion is measurable. A similar, but less pronounced, effect occurs in the Ni–Al–W system (T7 and T8), reflecting the higher diffusivity of W relative to Re. At the opposite extreme, the shallowest serpentine paths are observed in the Ni–Al–Ta system (T5 and T6), consistent with the relatively high diffusivity of Ta. Overall, the depth of serpentine deviation varies in the sequence Ni–Al–Ta < Ni–Al–Cr < Ni–Al–Mo < Ni–Al–W < Ni–Al–Re, reflecting the decreasing diffusivity of X (Ta, Cr, Mo, W, Re), as corroborated by the main interdiffusion coefficients presented in Fig. 10.

It is important to note that while diffusion coefficients are sensitive to diffusion length, the Gibbs triangle paths reflect only compositional variation, independent of length. The agreement between the diffusion coefficients derived from composition–distance profiles and the observed serpentine trajectories confirms the reliability of the data and provides a consistent explanation for the diffusion behaviour across different Ni–Al–X systems.

#### *3.2.2 Physics-informed neural network (PINN)-based numerical optimisation for the extraction of composition-dependent diffusion coefficients*

As discussed earlier, diffusion coefficients at a given composition can be determined either by intersecting two diffusion paths or by evaluating them at the Kirkendall marker plane. To obtain composition-dependent diffusion coefficients across the entire compositional range traversed by the diffusion paths, a physics-informed neural network (PINN)–based inverse optimization method is employed. This approach optimizes the diffusion profiles while rigorously enforcing the governing diffusion equations [28]. Historically, various numerical inverse methods have been applied to optimize single diffusion profiles, primarily for extracting averaged coefficients such as $\widetilde{D}_{ij}^{n}$ [22, 26, 79-86]. However, in the absence of experimentally determined diffusion coefficients used as equality constraints, the resulting composition-dependent coefficients cannot be uniquely determined and may lack physical reliability, as demonstrated in the following analysis.

We first examine the optimization of a single diffusion profile for extracting composition-dependent interdiffusion coefficients, $\tilde{D}_{ij}^{n}$, a practice commonly employed in earlier studies irrespective of the numerical method adopted. As an illustrative case, the diffusion profile obtained from the Ni/Ni5Cr5Al diffusion couple annealed at 1100 °C for 50 h is considered. As shown in Fig. 14a, the optimized profile reproduces the smoothed experimental data with high fidelity; however, the extracted $\tilde{D}_{ij}^{n}$ values deviate substantially from the experimentally determined coefficients. In the figure, the optimized coefficients are represented by solid lines, whereas the experimental estimates are indicated by filled triangles. Notably, the experimental results identify $\tilde{D}_{\mathrm{CrAl}}^{\mathrm{Ni}}$ as the smallest among the four coefficients, while unconstrained optimization yields it as the largest. Furthermore, the optimized main interdiffusion coefficients for Al and Cr differ by more than an order of magnitude, contradicting the trends inferred from experimental estimation. This inconsistency is resolved when the experimentally estimated $\tilde{D}_{ij}^{n}$ at the composition intersection are imposed as equality constraints, as shown in Fig. 14b. Under these conditions, the optimized coefficients converge with the experimental values while preserving excellent agreement between the experimental and optimized diffusion profiles.

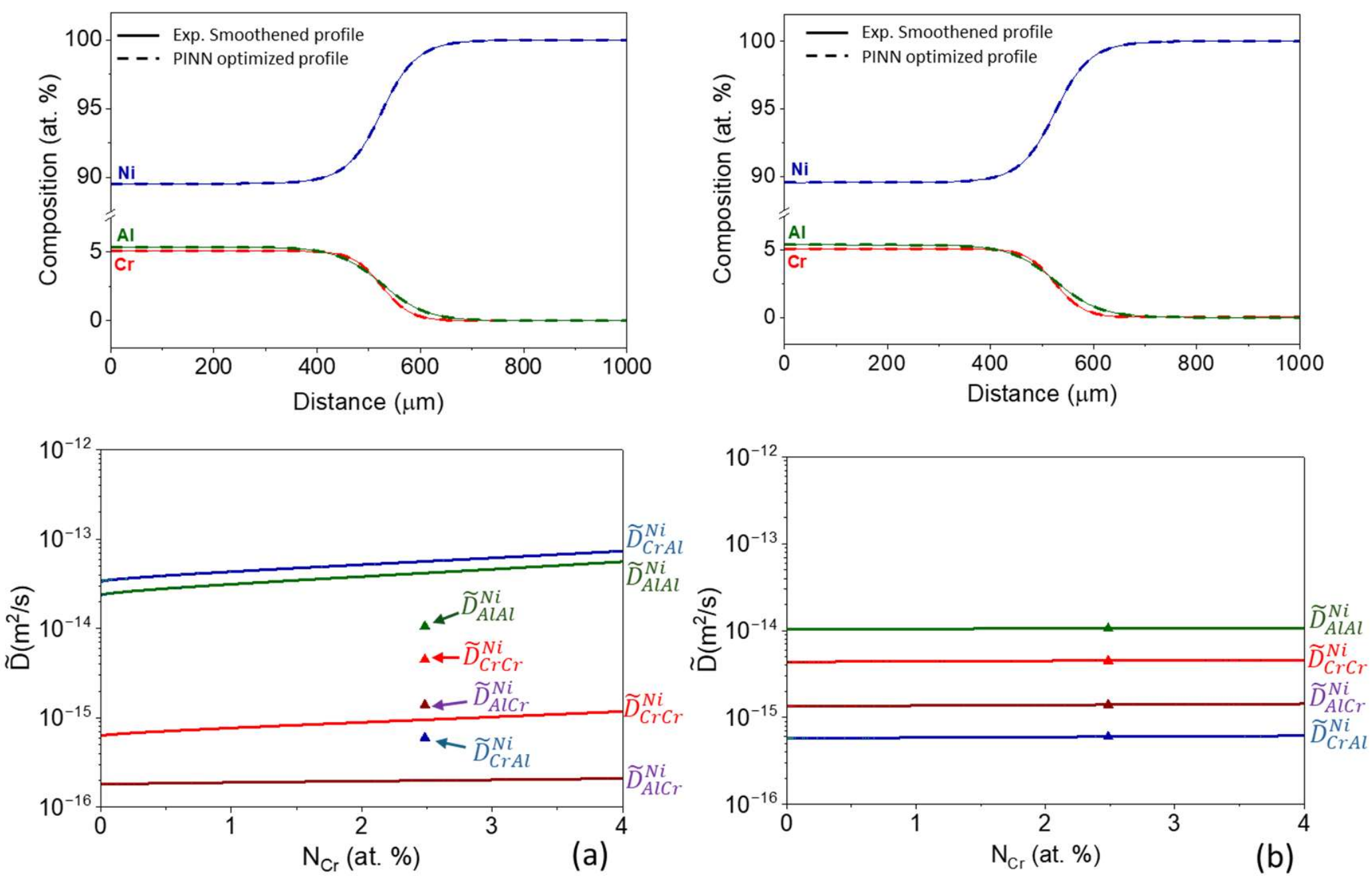

Fig. 14 Single diffusion-profile optimisation for extracting composition-dependent $\widetilde{D}$ from the Ni/Ni5Cr5Al diffusion couple (1100 °C, 50 h). Top: comparison between the smoothed experimental profiles (solid) and PINN-optimised profiles (dashed). Bottom: extracted interdiffusion coefficients $\widetilde{D}_{ij}^{Ni}$ as functions of $N_{Cr}$. (a) Optimisation without equality constraints, (b) optimisation with equality constraints imposed at the intersecting composition (filled triangles indicate experimentally estimated values).

Composition-dependent interdiffusion coefficients, $\widetilde{D}_{ij}^{n}$, can be directly estimated from two intersecting diffusion profiles. The central question, however, is whether the same pair of profiles can be employed in a combined optimization to yield reliable $\widetilde{D}_{ij}^{n}$ values without imposing experimentally determined data as equality constraints. The outcome is presented in Fig. 15, where excellent agreement is observed between the experimental and optimized diffusion profiles for both diffusion couples. Despite this, unconstrained optimization (i.e., optimization performed without equality constraints) fails to produce physically consistent $\widetilde{D}_{ij}^{n}$. Specifically, while experimental estimation identifies the cross interdiffusion coefficient $\widetilde{D}_{\mathrm{CrAl}}^{\mathrm{Ni}}$ as the lowest among the four coefficients, unconstrained optimization predicts it to be among the highest, comparable to $\widetilde{D}_{\mathrm{AlAl}}^{\mathrm{Ni}}$. Thus, the same limitation persists whether optimization is conducted using a single diffusion profile or two intersecting profiles. This discrepancy arises because the direct experimental method provides a well-defined estimate at the intersection composition, whereas profile-based inverse optimization admits multiple admissible solutions. Since each $\widetilde{D}_{ij}^{n}$ is parameterized using four optimization parameters (see Eq. 10), the optimizer may converge to different parameter sets that reproduce the diffusion profiles but yield non-unique $\widetilde{D}_{ij}^{n}$.

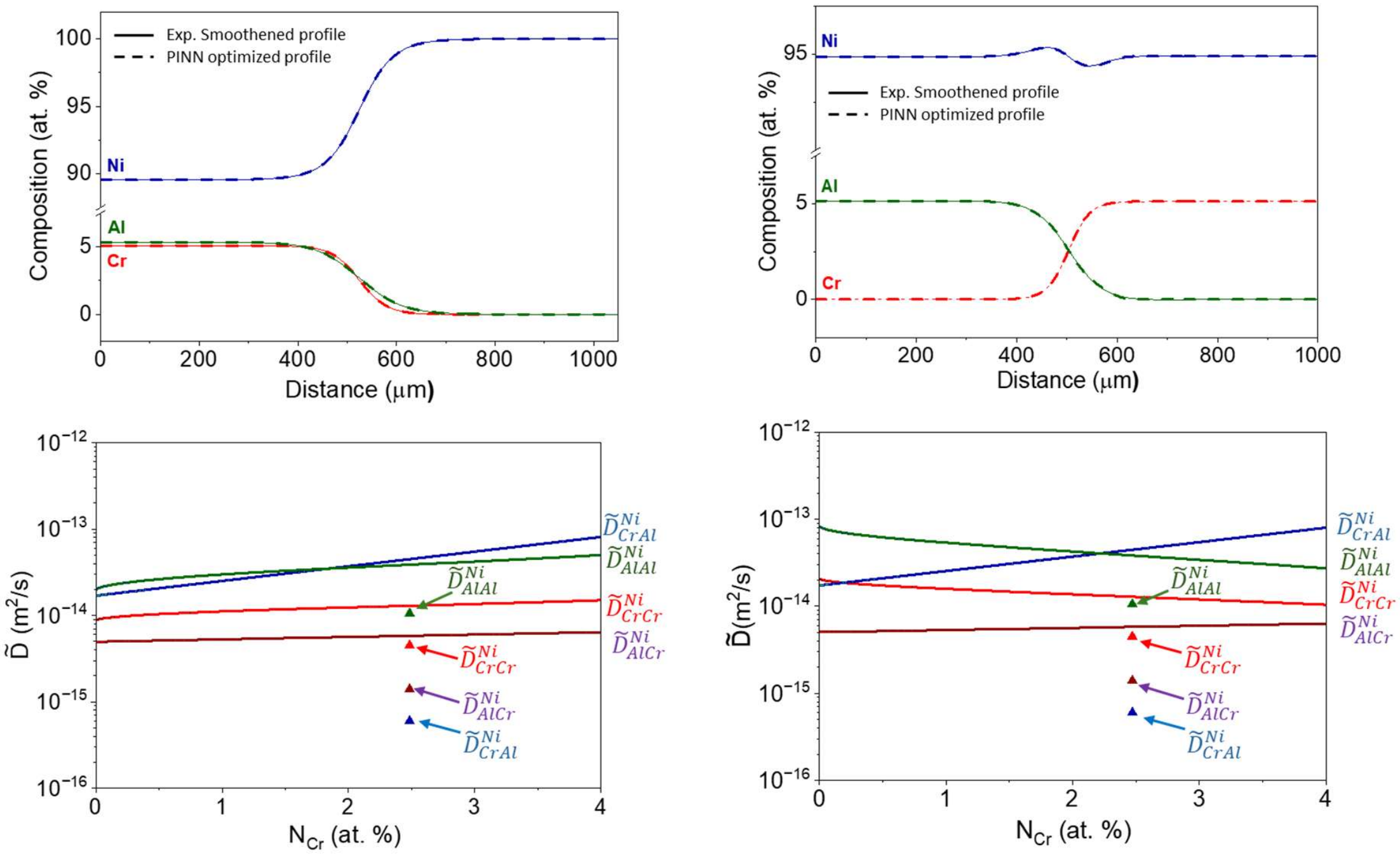

Fig. 15 A combined PINN optimisation is performed using diffusion profiles from two diffusion couples, Ni/Ni5Cr5Al and Ni5Al/Ni5Cr, annealed at 1100 °C for 50 h. Without equality constraints from experimentally estimated $\widetilde{D}_{ij}^{n}$ the extracted $\widetilde{D}_{ij}^{n}$ do not match the experimentally estimated data.

As a subsequent step, the $\widetilde{D}_{ij}^{n}$ values estimated at the intersecting composition are imposed as equality constraints. As shown in Fig. 16, the constrained optimization preserves the excellent agreement between the smoothed experimental and optimized diffusion profiles for both diffusion couples, while simultaneously recovering the correct relative magnitudes of the $\widetilde{D}_{ij}^{n}$. Because the diffusion couples examined here span relatively narrow compositional ranges, imposing equality constraints at a single composition is sufficient to suppress non-physical drift away from the intersection composition. Nevertheless, for achieving robust outcomes across broader compositional ranges, experimental data at multiple compositions are essential to capture the correct variation of $\widetilde{D}_{ij}^{n}$ with composition.

In addition to interdiffusion-coefficient estimates at the intersecting composition, impurity diffusion coefficients ($D_{i(Ni)}^{imp}$) extracted from binary Ni–X diffusion couple experiments can provide useful constraints, but they cannot be imposed at the pure-Ni end member because no diffusion profile (and hence no composition gradient) exists there. Instead, $D_{i(Ni)}^{imp}$ are applied at small but finite solute contents (e.g., 0.25 at.% X and 0.25 at.% Al), where a meaningful diffusion profile is present. At such compositions, the corresponding $\widetilde{D}_{ij}^{n}$ can be computed using thermodynamic factors and then enforced as equality constraints. Since $D_i^*$ are also estimated at the intersecting composition in the present study, a practical strategy is to combine $D_{i(Ni)}^{imp}$-based constraints at low solute contents with $D_i^*$-based constraints at the intersecting composition, thereby anchoring the optimisation at two distinct compositions and improving the robustness of the extracted composition-dependent diffusion coefficients. The single-profile optimization incorporating tracer and impurity diffusion coefficients is provided in the supplementary file. However, since intersecting diffusion profiles are available (with the exception of Ni–Al–Re), combined profile optimization was employed to extract data across the compositional range spanned by both diffusion profiles.

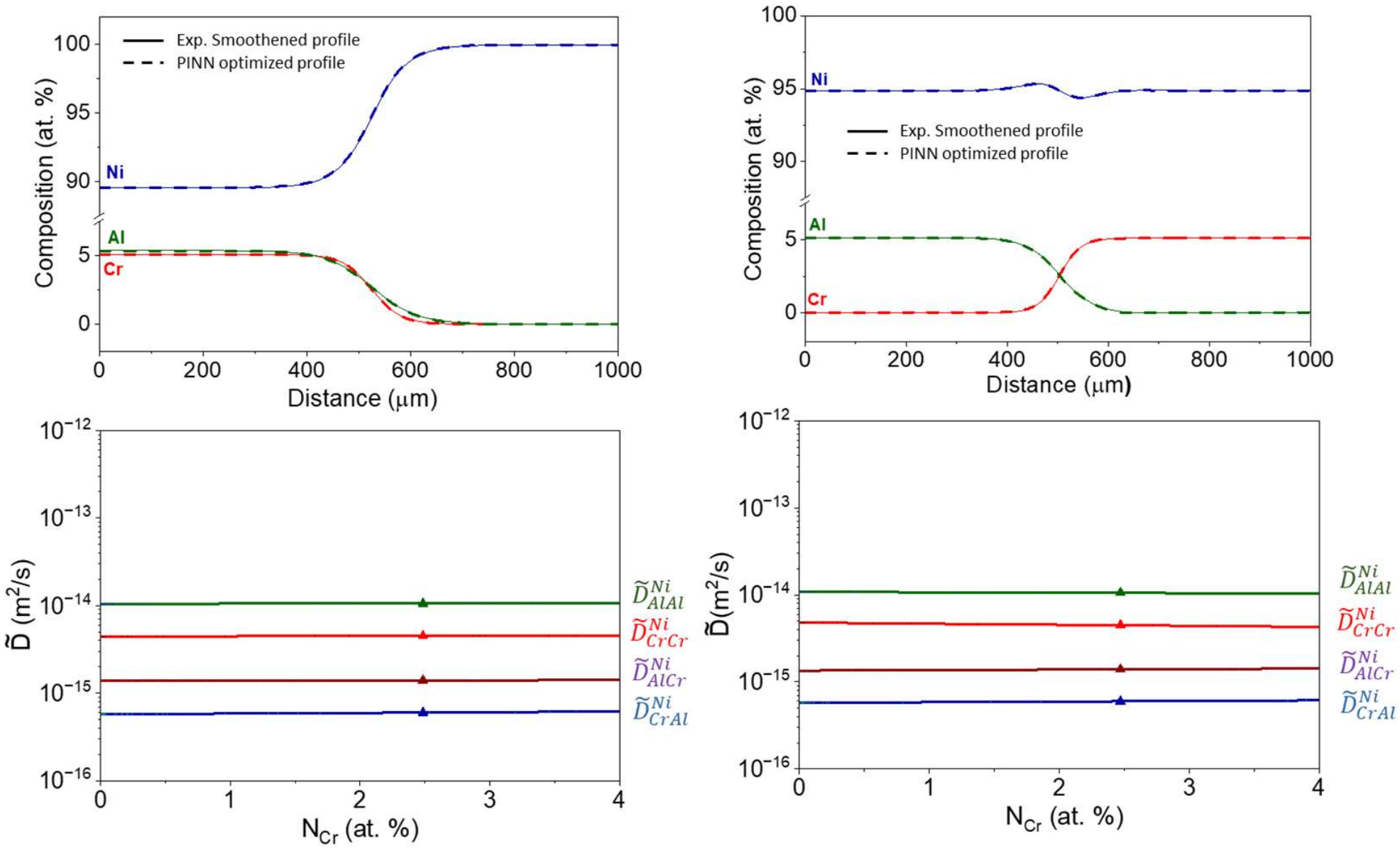

Fig. 16 A combined PINN optimisation of diffusion profiles produced by coupling (a) Ni/Ni5Cr5Al and (b) Ni5Al/Ni5Cr at 1100 °C for 50 h and imposing experimentally estimated values at the intersecting composition.

Initially, the combined diffusion-profile optimization was carried out without imposing experimentally determined data as equality constraints. It is important to emphasize that a physically consistent solution must satisfy both the governing diffusion equations and the experimentally estimated tracer and impurity diffusivities when these are applied as equality constraints. In the absence of such constraints, fitting only the diffusion profiles allows multiple parameter sets to reproduce the observed data, since each $D_i^*$ is expressed in terms of four optimization parameters (Eq. 3). This introduces ambiguity, and the optimizer may converge to a solution that matches the profiles but deviates from the experimentally estimated diffusivities, as illustrated in Fig. 17. Similar to single-profile optimization, the fitted values of $D_{\mathrm{Ni}}^*$ exhibit the trend $D_{\mathrm{Ni}}^* > D_{\mathrm{Cr}}^* \approx D_{\mathrm{Al}}^*$, whereas the experimentally determined trend is $D_{\mathrm{Al}}^* > D_{Cr}^* > D_{Ni}^*$. Thus, even an excellent agreement between optimized and smoothed experimental profiles does not necessarily ensure reliability.

To address this limitation, impurity diffusion coefficients from the Ni/Ni5Cr5Al profile, together with the $D_i^*$ values at the intersection of both diffusion profiles, were imposed as equality constraints in the combined optimization. As shown in Fig. 18, this approach enables robust extraction of $D_i^*$ across the composition range spanned by both profiles, yielding excellent agreement with the experimentally estimated tracer and impurity diffusion coefficients.

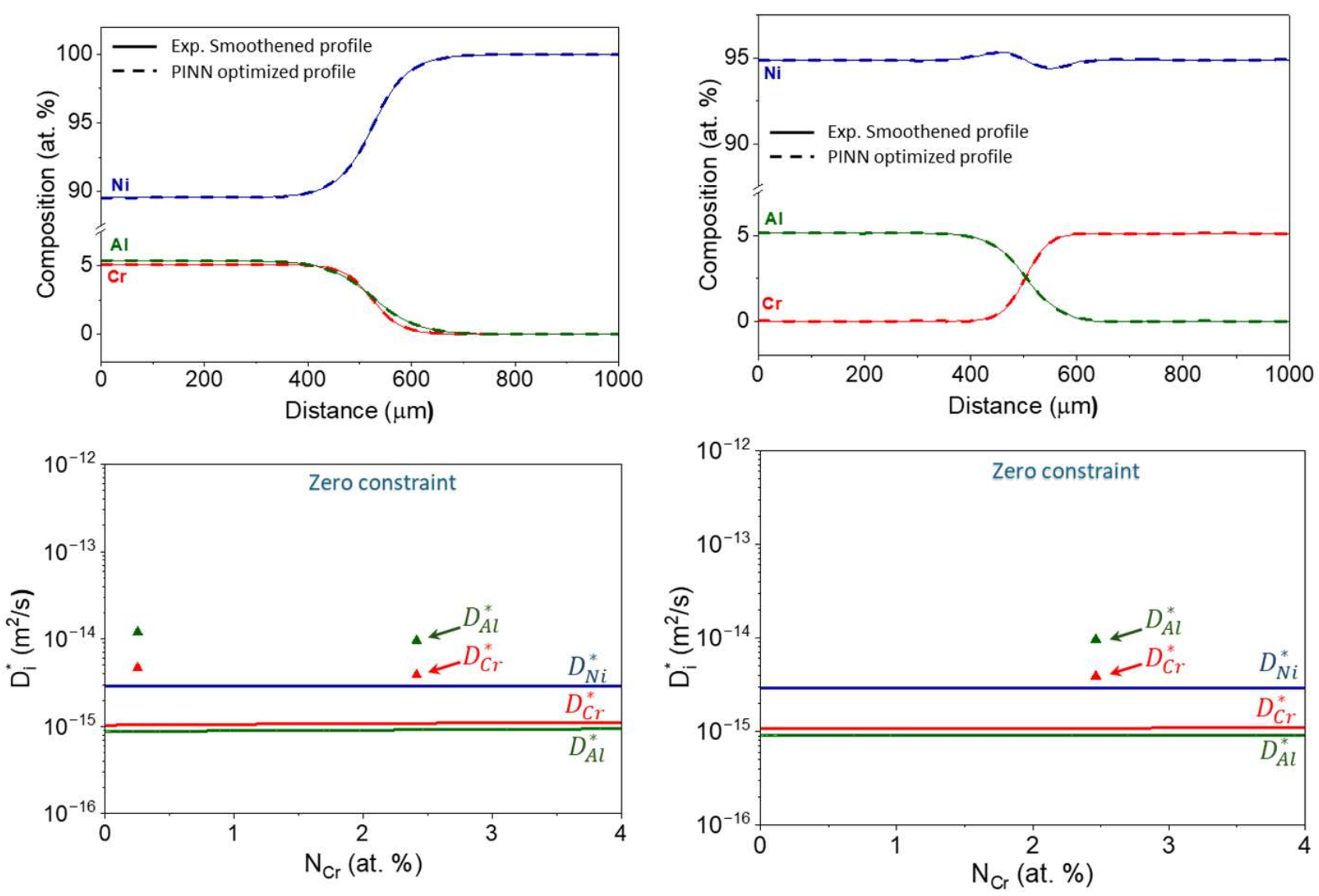

Fig. 17 Combined PINN optimisation of diffusion profiles produced by coupling Ni/Ni5Cr5Al and Ni5Al/Ni5Cr at 1100 °C for 50 h, performed without equality constraints from experimentally estimated tracer or impurity diffusion coefficients. Although the optimised profiles match the experimental profiles well, the extracted $D_i^*$ do not match the experimentally estimated values.

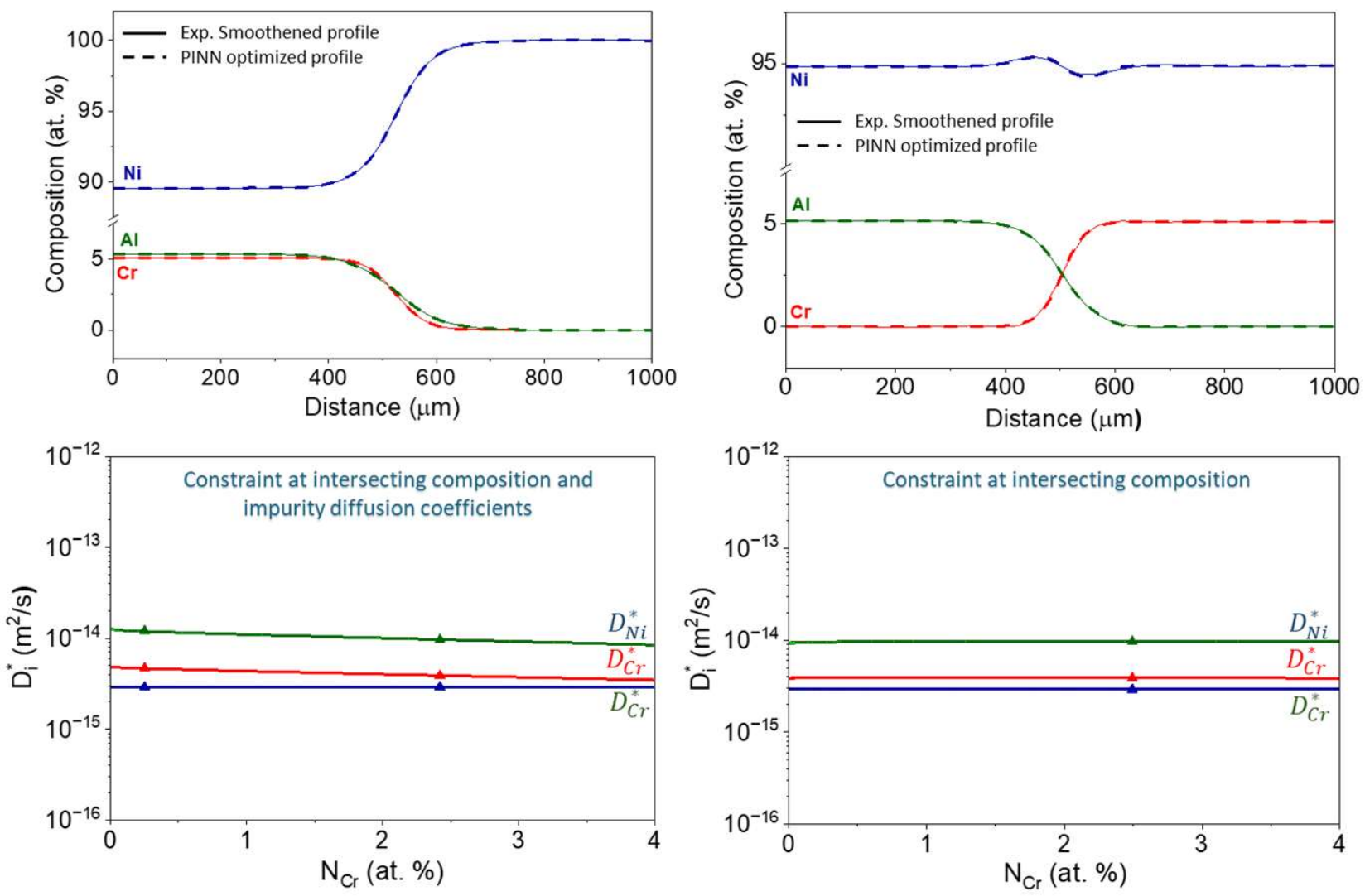

Fig. 18 Combined PINN optimisation of diffusion profiles produced by coupling Ni/Ni5Cr5Al and Ni5Al/Ni5Cr at 1100 °C for 50 h, with experimentally estimated impurity and $D_i^*$ at the intersecting composition imposed as equality constraints. The extracted $D_i^*$ show good agreement with the experimentally estimated tracer and impurity diffusion coefficients while maintaining an excellent match between the experimental and optimised diffusion profiles.

Based on the extracted optimisation parameters, the $D_i^*$ maps of the elements are plotted in Fig. 19. The diffusion paths of the diffusion profiles used for this optimisation are superimposed to visualise the change in diffusion coefficients along these paths. Along the diffusion path produced by coupling Ni and Ni5Cr5Al, both $D_{Cr}^*$ and $D_{Al}^*$ show a slight decrease with increasing solute content at 1100 °C. On the other hand, there is not much difference along the diffusion path of the diffusion profile produced by coupling Ni5Cr and Ni5Al. The variation is small due to the relatively narrow composition range considered. Moreover, extending the composition range to the solid-solution limit does not qualitatively change the trends observed here, as indicated by our unpublished Ni–Al–Ta analysis over a wider composition window in this phase. The optimisation parameters calculated for Ni–Al–X (X = Cr, Mo, Ta, W) systems, following a similar line of treatment, are listed in the supplementary file.

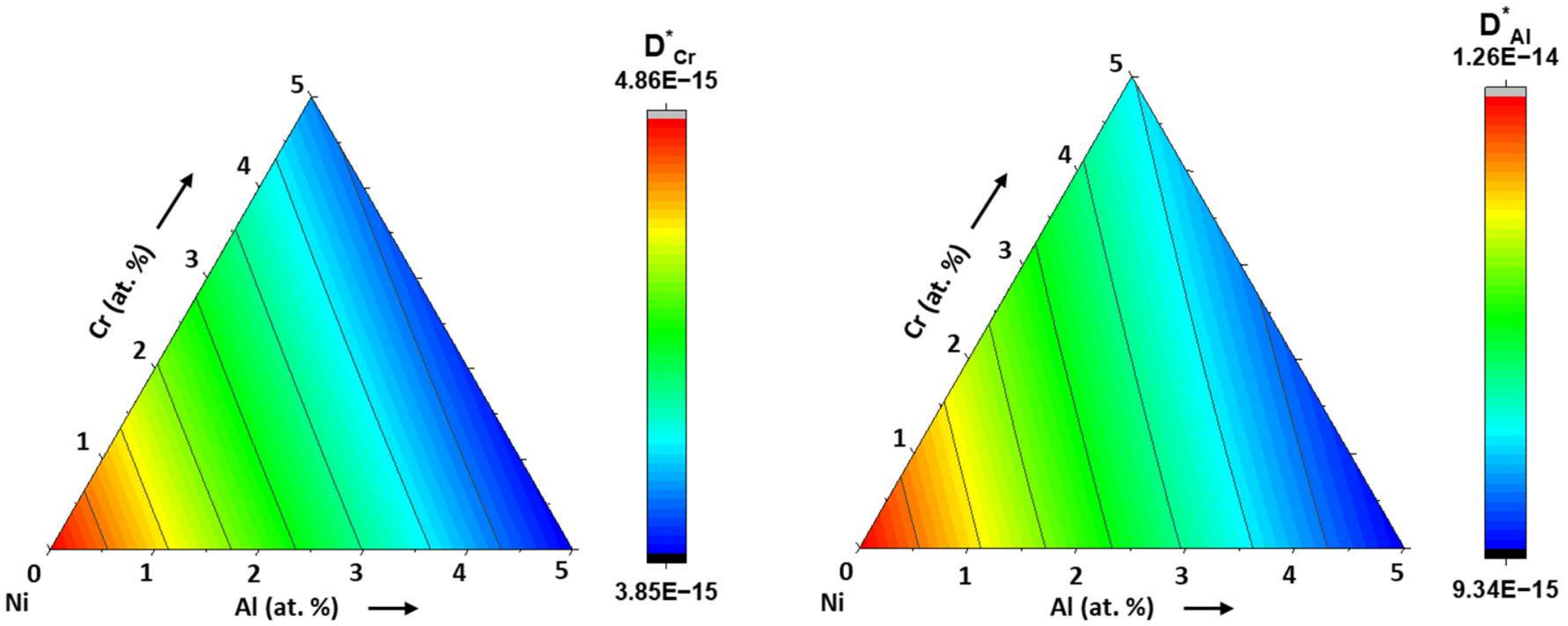

Fig. 19 $D_i^*$ maps across the composition range over the diffusion couple in the Ni-Cr-Al system, 1100 ºC, (a) Cr and (b) Al.

The transferability of the optimized parameters can be evaluated by calculating $\widetilde{D}$ in binary Ni–X and Ni–Al systems, and comparing these values with experimentally determined $\widetilde{D}$. Using the parameters obtained from the ternary optimization, the corresponding $D_i^*$ in the binary limits are derived from the reduced forms of the composition-dependent parameterization. Specifically, the expressions for $D_{\mathrm{Ni}}^*$ in the binary limits are given as follows:

For the Ni-Cr binary system:

$$D_{Ni}^* = D_o \exp\left(\theta_0^{Ni} + \theta_{Cr}^{1,Ni} N_{Cr}\right) \tag{25a}$$

$$D_{Cr}^* = D_o \exp\left(\theta_0^{Cr} + \theta_{Cr}^{1,Cr} N_{Cr}\right) \tag{25b}$$

For the Ni-Al binary system:

$$D_{Ni}^* = D_0 exp\left(\theta_0^{Ni} + \theta_{Cr}^{1,Ni} N_{Cr}\right) \tag{26a}$$

$$D_{Al}^* = D_0 exp\left(\theta_0^{Al} + \theta_{Al}^{1,Al} N_{Al}\right) \tag{26b}$$

As already mentioned, $D_0 = 2.77 \times 10^{-16}\ m^2/s$ is used in this study for all the optimisation.

The intrinsic diffusion coefficients were calculated using the relation $D_i = D_i^*\ \varphi$, where $\varphi = \left(\frac{\partial \ln\ a_i}{\partial \ln\ N_i}\right)$. The corresponding $\widetilde{D}$values in a binary $i$–$j$ system were then obtained from $\widetilde{D} = N_j D_i + N_i D_j$[92]. The binary $\widetilde{D}$ values predicted from the ternary-optimized parameters were compared with experimentally determined data, as shown in Fig. 20. Excellent agreement was achieved for the Ni–Cr system, while good agreement was also observed for the Ni–Al system when compared with literature values [18]. These findings demonstrate that parameters optimized in ternary systems can be reliably extended to the corresponding binary limits. In cases where larger discrepancies arise, binary $D_i^*$ data may be incorporated as additional equality constraints; however, such adjustments were not necessary in the present study.

The two-profile combined optimization strategy was applied to the remaining ternary systems, with the exception of Ni–Al–Re. In this case, diffusion coefficients were determined from a single diffusion profile by utilizing the Kirkendall marker plane.

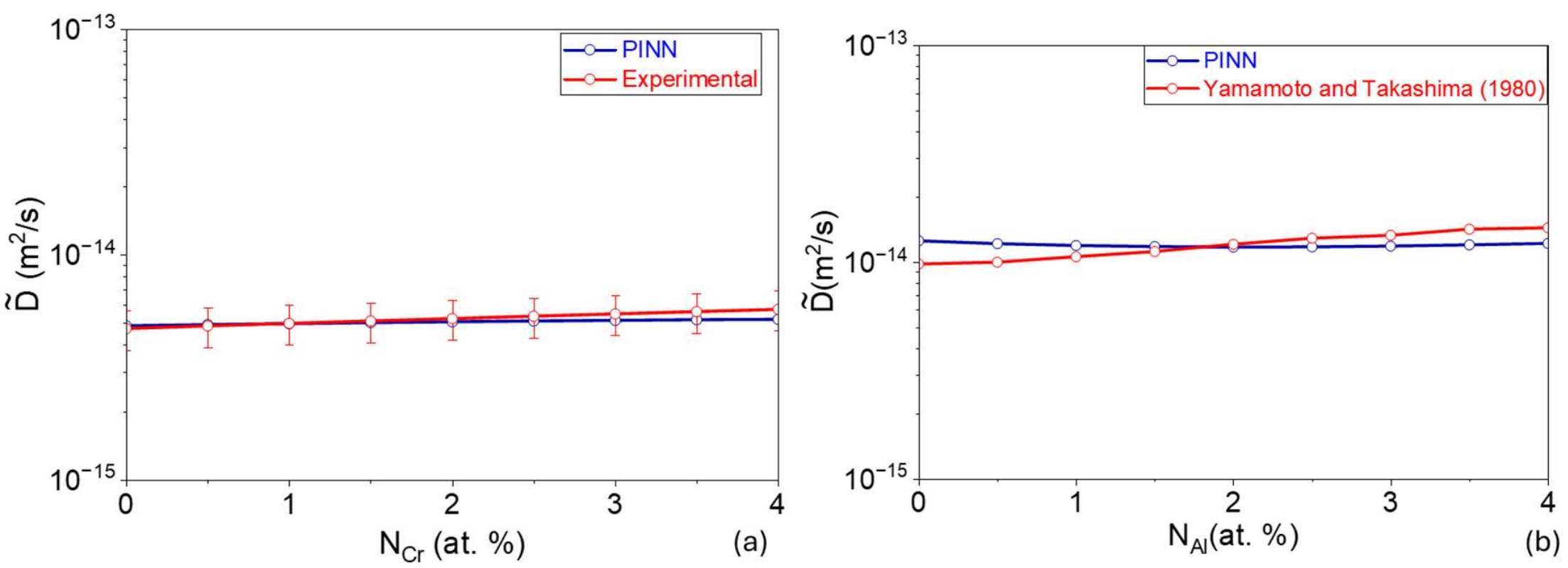

Fig. 20 Comparison between binary $\widetilde{D}$ predicted from ternary-optimised parameters and experimentally estimated values for (a) Ni–Cr and (b) Ni–Al systems.

In the Ni–Al–Re system, diffusion coefficients can be directly determined only at a single composition using the Kirkendall marker plane. However, by imposing equality constraints at more than one compositions, it becomes possible to reliably extract composition-dependent $D_{\mathrm{i}}^*$. In the present approach, the tracer diffusivity of Re at the Re-rich end (≈5 at.% Re) was obtained from the binary Ni–Re system by relating the binary $\widetilde{D}$ to the intrinsic diffusivity, expressed as $\widetilde{D}(\mathrm{Re}) \approx D_{\mathrm{Re}} = D_{\mathrm{Re}}^* \emptyset$, where $\varphi_{\mathrm{Re}} = \left(\frac{\partial \ln a_{\mathrm{Re}}}{\partial \ln N_{\mathrm{Re}}}\right)$ is the thermodynamic factor under the approximation employed here. The tracer diffusivity of Al on the Ni–Re alloy side was determined using the Hall method [29,87,88], applied to impurity diffusion coefficients within the ternary diffusion profile (see supplementary information for details of this calculation). These tracer and impurity estimates were then imposed as equality constraints at the Re-rich side of the diffusion couple, together with constraints at the Kirkendall marker plane, as illustrated in Fig. 21. The optimized parameters for all temperatures and systems are provided in the supplementary file.

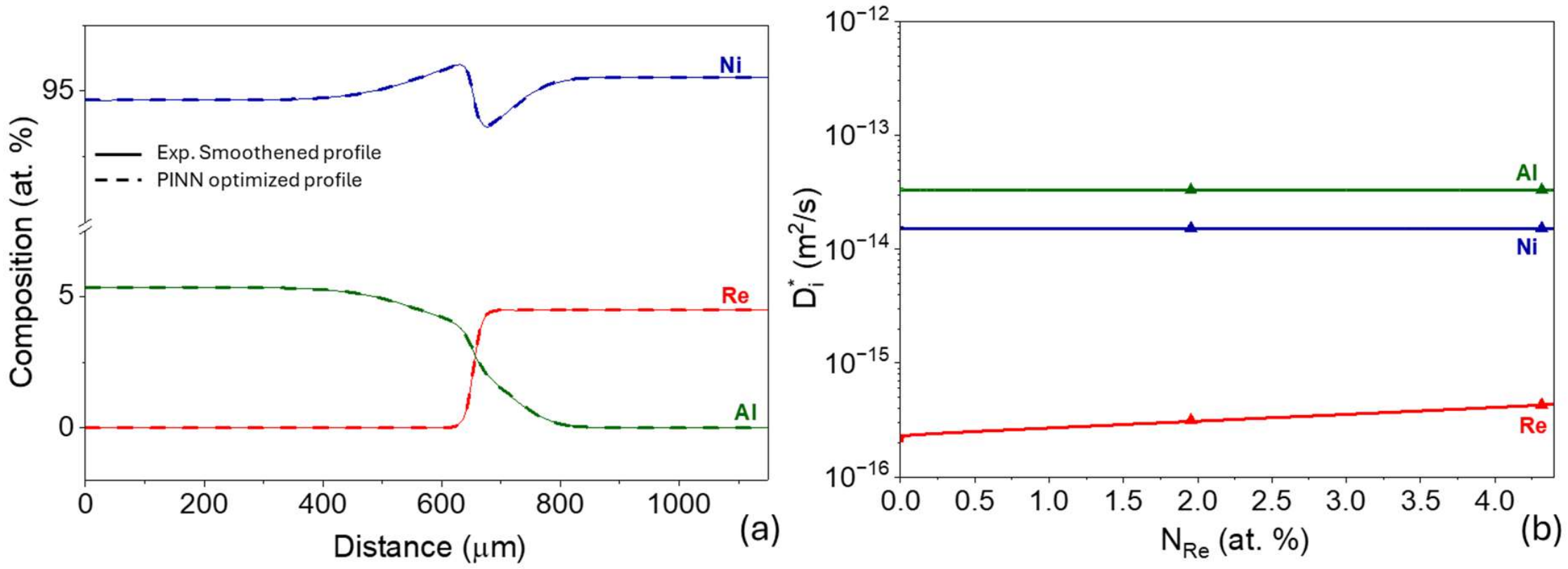

Fig. 21 PINN optimisation of a ternary Ni–Al–Re diffusion profile at 1200 °C for 50 h, using equality constraints at the Re-rich side and at the Kirkendall marker plane. (a) Comparison between smoothed experimental and optimised composition profiles. (b) extracted $D^*$ as a function of Re atomic fraction (symbols indicate constrained values).

## 4. Conclusion

This study constitutes the first comprehensive investigation of temperature-dependent diffusion coefficients in comparison to binary Ni–X (X = Cr, Mo, Ta, W, Re, Al) and ternary Ni–Al–X (X = Cr, Mo, Ta, W, Re) systems. In the binary alloys, $\widetilde{D}$ closely approximate the $D_X$ in these X-lean compositions, thereby reflecting the relative diffusion rates of alloying elements in Ni. The observed diffusivity sequence is $D_{Al} > D_{Ta} \approx D_{Cr} > D_{Mo} > D_W > D_{Re}$. Even the impurity diffusion coefficient of X in Ni, $D_{\mathrm{X(Ni)}}^{\mathrm{imp}}$, obtained by extrapolating $\widetilde{D}$ to zero atomic percentage of X, follows the same trend due to linear variation with composition. Among these, Ta exhibits the lowest diffusion activation energy, whereas Re shows the highest. First-principles calculations corroborate the observed trend, attributing the differences predominantly to variations in migration energies. Among the investigated elements, Ta exhibits the lowest migration energy, whereas Re possesses the highest.

In most Ni–Al–X systems, diffusion coefficients could be estimated by intersecting diffusion paths, a conventional approach, except in Ni–Al–Re. In this system, the intersecting composition lies close to the end-member alloys, introducing large uncertainties in composition gradients and, consequently, in the diffusion coefficients. To mitigate this, a single-profile estimation at the Kirkendall marker is employed. While the main interdiffusion and intrinsic diffusion coefficients of X remain comparable to those in binary systems, cross-interdiffusion coefficients exert a significant influence by either enhancing or reducing flux depending on their sign and the diffusion direction of the elements. This effect is illustrated by comparing the relative diffusion lengths in diffusion profiles from two couples: one in which X and Al diffuse in the same direction, and the other in which they diffuse in opposite directions.

The distinction between $\tilde{D}_{ij}^{n}$ and $D_{ij}^{n}$ is particularly significant in the Ni–Al–Re system, even at low concentrations of Al and X. Specifically, $D_{ReAl}^{Ni}$ indicates only a negligible positive diffusional interaction between Re and Al, whereas $\tilde{D}_{\mathrm{ReAl}}^{n}$—defined as the average of three coefficients ($D_{ReAl}^{Ni}, D_{AlAl}^{Ni}, D_{NiAl}^{Ni}$)—erroneously suggests a pronounced negative interaction. Exclusive reliance on $\tilde{D}_{ij}^{n}$, as is often the case in practice, can therefore lead to misleading interpretations. While $\tilde{D}_{ij}^{n}$ values are useful for correlating directly with diffusion profiles, discussions of element-specific diffusional interactions should, wherever possible, be based on $D_{ij}^{n}$, since these parameters more accurately capture the intrinsic properties of individual species. Furthermore, the estimated diffusion coefficients provide a coherent explanation for the serpentine trajectories observed in diffusion paths plotted on a Gibbs triangle, thereby validating the estimates and elucidating differences in path depth across systems.

Because diffusion data in ternary systems are available only at specific compositions—typically at intersecting diffusion paths or at the Kirkendall marker plane—a physics-informed neural network (PINN)-based inverse optimization framework is employed to extract composition-dependent diffusion coefficients across the full compositional range of diffusion couples. The inverse problem is formulated in Boltzmann coordinates and trained using a composite loss function incorporating physics residuals, boundary conditions, data misfit, and equality constraints. A central outcome of the PINN study is that optimization based solely on diffusion profiles does not guarantee physically reliable coefficients, even when the optimized profiles reproduce experimental data with high fidelity. Non-identifiability arises because each diffusion coefficient is parameterized by a set of trainable variables, allowing multiple admissible profile-fitting solutions that yield incorrect magnitudes of $\tilde{D}_{ij}^{n}$ or $D_{ij}^{n}$. This issue is demonstrated for both single-profile and combined two-profile optimizations.

Reliable extraction of composition-dependent diffusion coefficients is achieved only when independently estimated diffusion data are imposed as equality constraints. Specifically, enforcing experimentally determined $\tilde{D}_{ij}^{n}$ at intersecting compositions anchors interdiffusion-coefficient optimization, while imposing tracer or impurity diffusion coefficients at selected compositions anchors tracer-diffusion optimization. Applying constraints across multiple compositions enhances robustness and enables recovery of the correct composition dependence, particularly across broader compositional ranges.

Finally, the extendibility of optimization parameters derived from ternary diffusion couples is validated through binary-limit comparisons. Interdiffusion coefficients computed for binary Ni–Cr and Ni–Al alloys using ternary-optimized parameters show a very good agreement with experimentally estimated data in this study and available in literature, confirming that the learned parameterization captures physically meaningful trends and can be transferred to related compositional trajectories.

Acknowledgement: We acknowledge the use of EPMA in the Advanced facility for Microscopy and Micro Analysis (AFMM) at IISc. A. Paul and S. Bhattacharyya acknowledge the financial support from SERB, India (project No. CRG/2021/001842). A. Paul also acknowledges the financial support from J.C. Bose Grant, ANRF (project no. ANRF/JBG/2025/000033/EST). We acknowledge the computational resources provided by the Supercomputer Education and Research Centre (SERC), IISc. G.S. Gautam acknowledges the computing time provided to them on the high-performance computers Noctua 1 and Noctua 2 at the NHR Centre PC2. This was funded by the Federal Ministry of Education and Research and the state governments participating on the basis of the resolutions of the GWK for the national high-performance computing at universities (www.nhr-verein.de/unsere-partner). The computations for this research were performed using computing resources under the project hpc-prf-emdft.

**Supplementary**

## S1. Diffusion profile in the Ni-Al-Re system

Diffusion profile produced by coupling Ni with Ni-5Al-5Re at 1250 ºC is shown. Although possible, this is not utilised for the calculation of tracer diffusion coefficients at the Kirkendall marker plane because of a plateau in the Al diffusion profile, which may introduce higher errors than the diffusion profile produced by coupling Ni-5Al and Ni-5Re, as shown in Fig. 9.

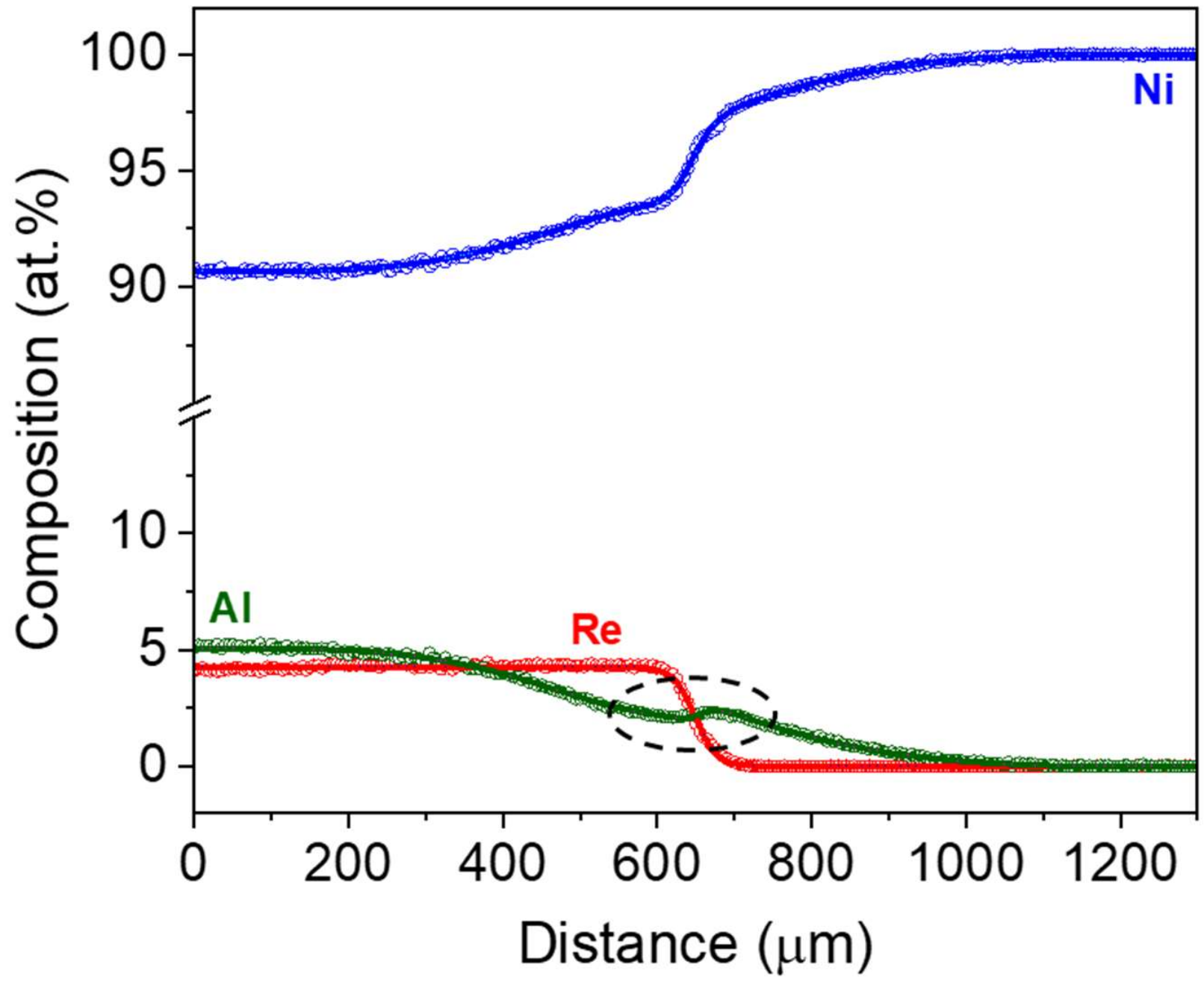

Fig. S1 A diffusion profile produced by coupling Ni and Ni-5Al-5Re at 1250 ºC after annealing for 50 h.

## S2 Comparison of activation energies calculated in binary Ni-X systems

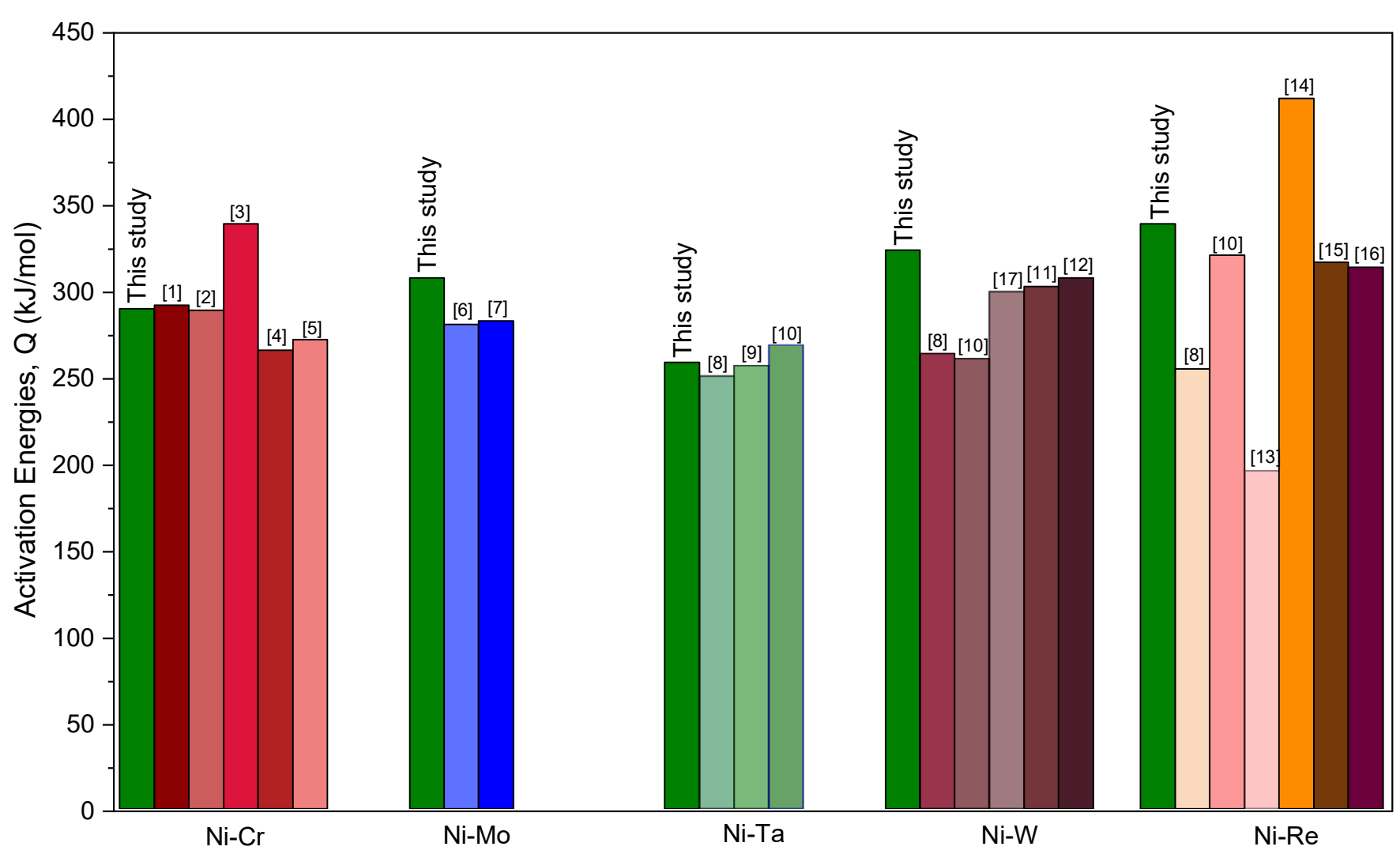

Fig S2 Activation energies calculated in this study compared with the data available in the literature [1-16] in the article.

## S3 Comparison between interdiffusion coefficients in binary and ternary systems

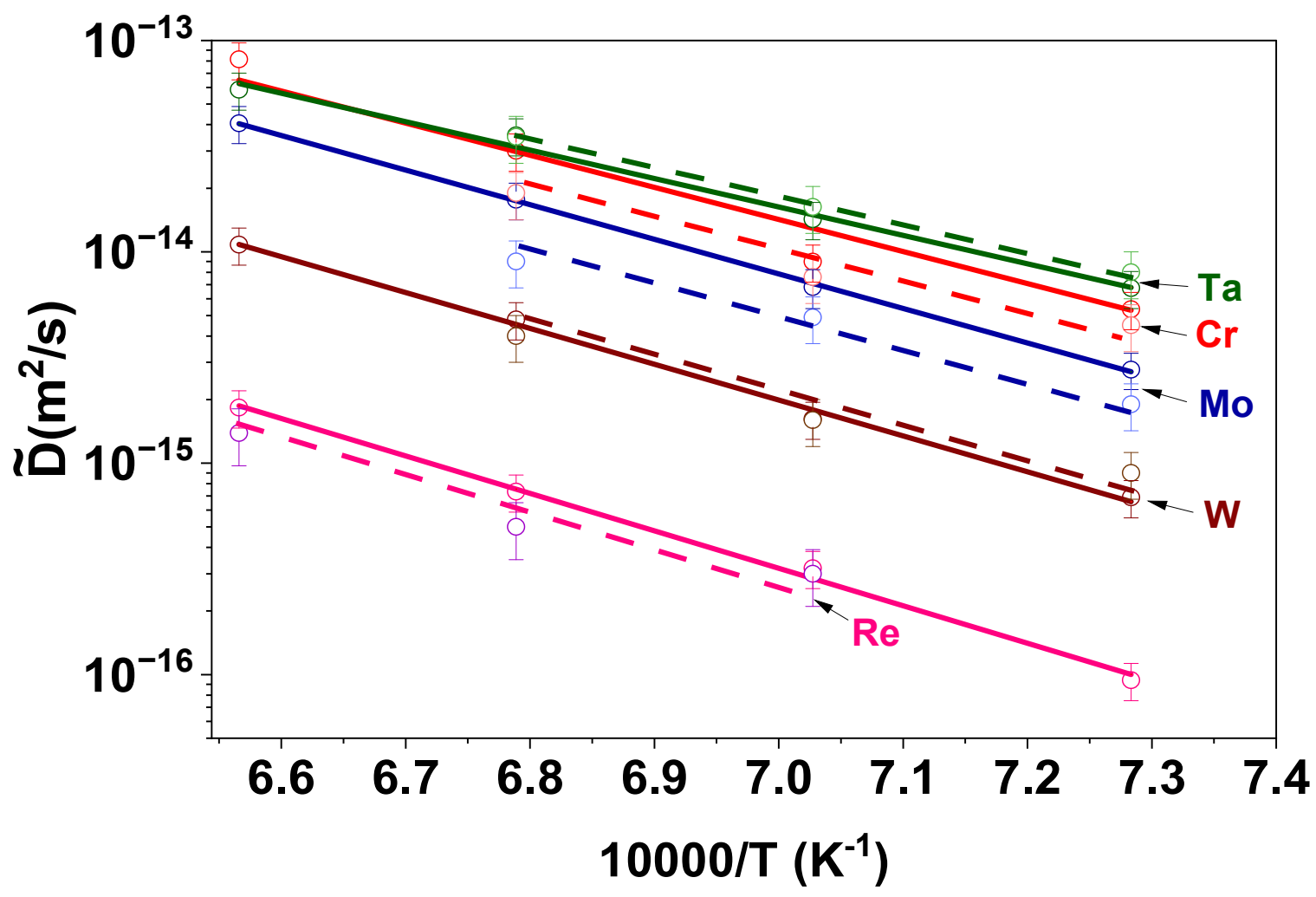

Fig. S3 Comparison of $\widetilde{D}$ of binary Ni-X and $\widetilde{D}_{XX}^{Ni}$ of ternary Ni-Al-X systems.

## S4 Thermodynamic factors utilized for calculations of tracer diffusion coefficients in the Ni-Al-X systems

Table S1: Thermodynamic factors calculated by extracting the activity data from TCNI9 of ThermoCalc (Note: a little difference in intersecting compositions at different temperatures may have resulted from differences in diffusion paths at different temperatures, or sometimes because of repetition of experiments with a very little difference in end-member composition of alloys melted in different batches).

| Temperature (°C) | Composition (at.%) | $\phi_{NiX}^{Ni}$ | $\phi_{NiAl}^{Ni}$ | $\phi_{XX}^{Ni}$ | $\phi_{XAl}^{Ni}$ | $\phi_{AlX}^{Ni}$ | $\phi_{AlAl}^{Ni}$ |
|---|---|---|---|---|---|---|---|
| 1100 | $NiCr_{2.5}Al_{2.6}$ | -0.04 | -0.04 | 1.17 | 0.22 | 0.20 | 1.13 |
| | $NiMo_{2.4}Al_{2.6}$ | -0.03 | -0.03 | 1.14 | 0.20 | 0.18 | 1.12 |
| | $NiTa_{2.1}Al_{2.4}$ | -0.05 | -0.04 | 1.76 | 0.42 | 0.34 | 1.10 |
| | $NiW_{4.4}Al_{1.6}$ | -0.10 | -0.03 | 1.93 | 0.25 | 0.63 | 1.06 |
| 1150 | $NiCr_{2.3}Al_{2.9}$ | -0.03 | -0.04 | 1.15 | 0.23 | 0.19 | 1.14 |
| | $NiMo_{2.4}Al_{2.9}$ | -0.03 | -0.04 | 1.13 | 0.22 | 0.18 | 1.14 |
| | $NiTa_{2.5}Al_{2.6}$ | -0.06 | -0.04 | 1.86 | 0.42 | 0.39 | 1.10 |
| | $NiW_{3.3}Al_{2.5}$ | -0.07 | -0.04 | 1.68 | 0.36 | 0.47 | 1.10 |
| | $NiRe_{2.0}Al_{3.0}$ | -0.03 | -0.05 | 1.13 | 0.30 | 0.19 | 1.20 |
| 1200 | $NiCr_{2.4}Al_{2.8}$ | -0.03 | -0.04 | 1.10 | 0.21 | 0.18 | 1.12 |
| | $NiMo_{2.4}Al_{2.8}$ | -0.03 | -0.04 | 1.13 | 0.20 | 0.17 | 1.12 |
| | $NiTa_{2.6}Al_{2.4}$ | -0.06 | -0.04 | 1.85 | 0.38 | 0.39 | 1.09 |
| | $NiW_{3.8}Al_{2.0}$ | -0.08 | -0.03 | 1.74 | 0.27 | 0.49 | 1.10 |
| | $NiRe_{2.0}Al_{3.0}$ | -0.03 | -0.05 | 1.13 | 0.30 | 0.19 | 1.20 |
| 1250 | $NiRe_{1.7}Al_{3.1}$ | -0.02 | -0.04 | 1.08 | 0.30 | 0.11 | 1.19 |

**S5 The pre-exponential factors and activation energies of tracer diffusion coefficients in Ni, Ni-X and Ni-X-Al systems**

The pre-exponential factors ($D_o$) and activation energies ($Q$) are listed in Table S2. The impurity diffusion coefficients, $D_{X(Ni)}^{imp}$ of elements X in Ni. The tracer diffusion coefficients of X at 2.5 at.%X in the binary system are calculated from $\widetilde{D}(X) = N_X D_{\mathrm{Ni}} + N_{\mathrm{Ni}} D_X \approx D_X \approx D_X^* \Phi$, since interdiffusion coefficients of X are close to the intrinsic diffusion coefficients of X at this relatively low composition of X. Data for tracer diffusion coefficients calculated at compositions close to 2.5 at.%X and 2.5 at.%Al in the ternary Ni-Al-X systems are also included. The activation energies for these diffusion coefficients are found to be similar within the experimental error of calculations.

Table S2 The pre-exponential factors and activation energies for impurity (in Ni) and tracer diffusion coefficients at 2.5 at.%X in the Ni-X binary system and 2.5 at.%X and 2.5 at.%Al in the ternary Ni-Al-X systems.

| System | $D^{imp}_{i(Ni)}/D^*_i$ (Composition in at.%) | $D_0$ ($\times 10^{-4} m^2/s$) | Q |
|---|---|---|---|
| Ni/Ni-Cr/Ni-Al-Cr | $D^{imp}_{Cr(Ni)}$ | 5.5 (± 6) | 290 ± 15 |
| | $D^*_{Cr}$ (Ni2.5Cr) | 5.2 (± 0.6) | |
| | $D^*_{Cr}$ (Ni2.4Cr2.8Al) | 3.7 (± 0.4) | |
| Ni/Ni-Mo/Ni-Al-Mo | $D^{imp}_{Mo(Ni)}$ | 14.5 (± 1.4) | 308 ± 16 |
| | $D^*_{Mo}$ (Ni2.5Mo) | 12.5 (± 2.3) | |
| | $D^*_{Mo}$ (Ni2.4Mo2.8Al) | 8.7 (± 0.9) | |
| Ni/Ni-Ta/Ni-Al-Ta | $D^{imp}_{Ta(Ni)}$ | 0.5 (± 0.5) | 259 ± 13 |
| | $D^*_{Ta}$ (Ni2.5Ta) | 0.2 (± 0.2) | |
| | $D^*_{Ta}$ (Ni2.6Ta2.4Al) | 0.3 (± 0.3) | |
| Ni/Ni-W/Ni-Al-W | $D^{imp}_{W(Ni)}$ | 14.7 (± 1.2) | 324 ± 16 |
| | $D^*_{W}$ (Ni2.5W) | 9.2 (± 0.9) | |
| | $D^*_{W}$ (Ni3.8W2.0Al) | 6.9 (± 0.7) | |
| Ni/Ni-Re/Ni-Al-Re | $D^{imp}_{Re(Ni)}$ | 8.1 (± 0.8) | 339 ± 17 |
| | $D^*_{Re}$ (Ni2.5Re) | 7.5 (± 0.7) | |
| | $D^*_{Re}$ (Ni2Re3Al) | 3.4 (± 0.4) | |
| Ni/Ni-Al/Ni-Al-X (X = Cr, Mo, Ta, W, Re) | $D^{imp}_{Al(Ni)}$ | 1.9 (± 0.2) | 268 ± 13 |
| | $D^*_{Al}$ (Ni2.5Al) | 1.8 (± 0.3) | |
| | $D^*_{Al}$ (Ni2.4Cr2.8Al) | 1.5 (± 0.15) | |
| | $D^*_{Al}$ (Ni2.4Mo2.8Al) | 1.0 (± 0.1) | |
| | $D^*_{Al}$ (Ni2.6Ta2.4Al) | 1.9 (± 0.2) | |
| | $D^*_{Al}$ (Ni3.8W2.0Al) | 1.0 (± 0.1) | |
| | $D^*_{Al}$ (Ni2.0Re3.0Al) | 1.1 (± 0.1) | |

## S6 Single profile optimization with impurity and tracer diffusion coefficients

To optimize with impurity and tracer diffusion coefficients, we first examine a single diffusion profile obtained from the Ni/Ni5Cr5Al diffusion couple annealed at 1100 °C for 50 h (Fig. S4). As shown in Fig. S4a, the optimized profile reproduces the experimental diffusion profile with excellent fidelity; however, the extracted ($D_i^*$ deviate significantly from the experimentally determined values. As noted earlier, the Ni tracer diffusion coefficient is prescribed as the Ni self-diffusion coefficient across the composition range of interest, since in these Ni-rich alloys, the interdiffusion process is primarily governed by Cr and Al, and $D_{\mathrm{Ni}}^*$ cannot be reliably calculated from the diffusion profile alone. This prescribed $D_{\mathrm{Ni}}^*$ is then employed during optimization to extract the tracer diffusivities of Al and Cr. Experimentally, the ordering is found to be $D_{\mathrm{Al}}^* > D_{\mathrm{Cr}}^* > D_{\mathrm{Ni}}^*$, whereas profile-only optimization yields the opposite trend, $D_{\mathrm{Ni}}^* > D_{\mathrm{Cr}}^* \approx D_{\mathrm{Al}}^*$, despite the excellent agreement with the diffusion profile. This outcome clearly demonstrates that reproducing the diffusion profile alone does not ensure physically reliable tracer diffusivities, and that experimentally estimated values must be imposed as equality constraints to achieve robust extraction.

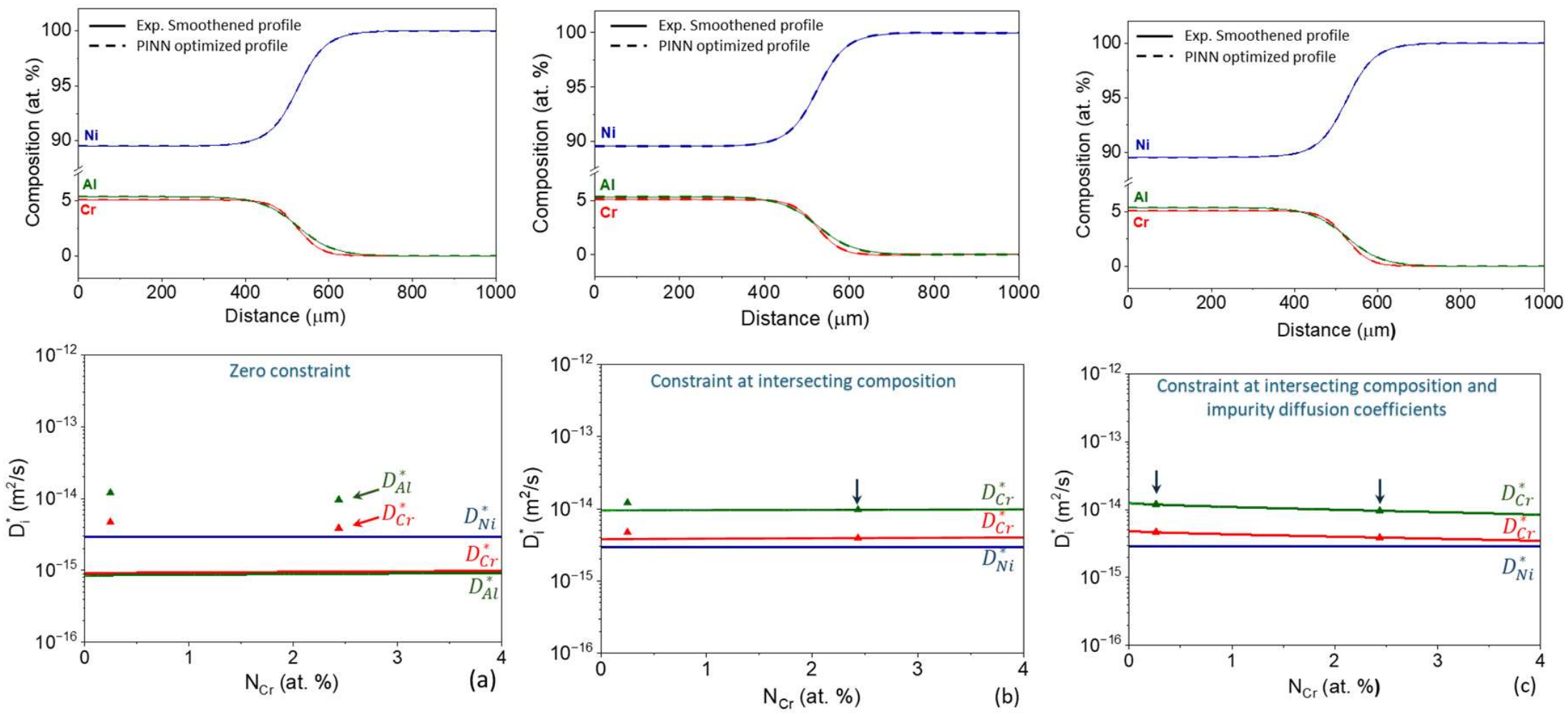

Fig. S4 PINN optimisation of the Ni/Ni5Cr5Al diffusion profile annealed at 1100 °C for 50 h, illustrating the role of tracer/impurity constraints. (a) Profile-only optimisation yields an excellent profile fit but incorrectly extracted $D^*$. (b) Imposing experimentally estimated $D_i^*$ at the constrained composition as equality constraints yields a reliable extraction while preserving the profile fit. (c) Adding impurity-diffusion constraints in addition to the tracer constraints further improves the extracted composition dependence of the $D^*$.

As a subsequent step, the experimentally estimated $D_i^*$ at the intersecting composition were imposed as equality constraints during optimization (indicated by the arrow in Fig. S4b). The constrained optimization preserves the excellent agreement between the experimental and optimized diffusion profiles, while simultaneously recovering the correct relative mobilities of the $D_i^*$. For enhanced robustness, it is generally advantageous to impose equality constraints across multiple compositions. Within the present framework, impurity diffusion coefficients inferred from binary Ni–X couples provide additional constraints; however, these cannot be applied at the pure-Ni end member because no diffusion profile—and hence no composition gradient—exists there. Instead, impurity-based constraints are imposed at a small but finite solute concentration (here 0.25 at.% X and 0.25 at.% Al), where tracer and impurity diffusivities are expected to be nearly equivalent. The resulting optimization outcome, shown in Fig. S4c, demonstrates improved consistency with experimentally estimated tracer and impurity diffusivities, while maintaining excellent agreement with the diffusion profile.

**S7: Calculation of impurity diffusion coefficients of Al from the ternary diffusion profile in the Ni-Al-Re system produced by coupling Ni-5Al and Ni-Re.**

The ternary diffusion profile produced in this system at 1200 ºC is shown in Fig. S5. The Hall method is used to calculate the Al impurity diffusion coefficient on the Ni-5Re side of the diffusion couple. Hall proposed a method for estimating the impurity diffusion coefficient of elements from the interdiffusion profile obtained using the diffusion couple method [S1, S2]. He proposed an analytical expression that can be used at very low concentrations via a probability plot of the measured composition profile. This can also be extended to the ternary or multicomponent systems [S3, S4]. The proposed relations in the Hall method are [S2]

$$\frac{N_i - N_i^-}{N_i^+ - N_i^-} = \frac{N_i}{N_i^+} = \frac{1}{2}(1 + erfU) \tag{S1}$$

Since $N_i^- = 0$ for the elements for which the impurity diffusion coefficients are estimated.

$$U = h\lambda + k, \tag{S2}$$

where, $\lambda = \frac{x - x_o}{t^{1/2}}$ is the Boltzmann parameter [S5-S6], $x_o$ is the initial contact plane of the diffusion couple or the Matano plane [S5]. In this, the composition vs. distance plot is converted to U vs. $\lambda$ linear plot to determine the values of $h$ (slope) and $k$ (intercept). Following, the impurity diffusion coefficient of the element is calculated from [S2]

$$D = \frac{1}{4h^2}\left[1 + k\sqrt{\pi}\exp{(U)^2}(1 + erfU)\right] \tag{S3}$$

The U vs. $\lambda$ plot for Al is shown in Figure S2, from which the impurity diffusion coefficient is calculated for PINN optimisation, as described in the article.

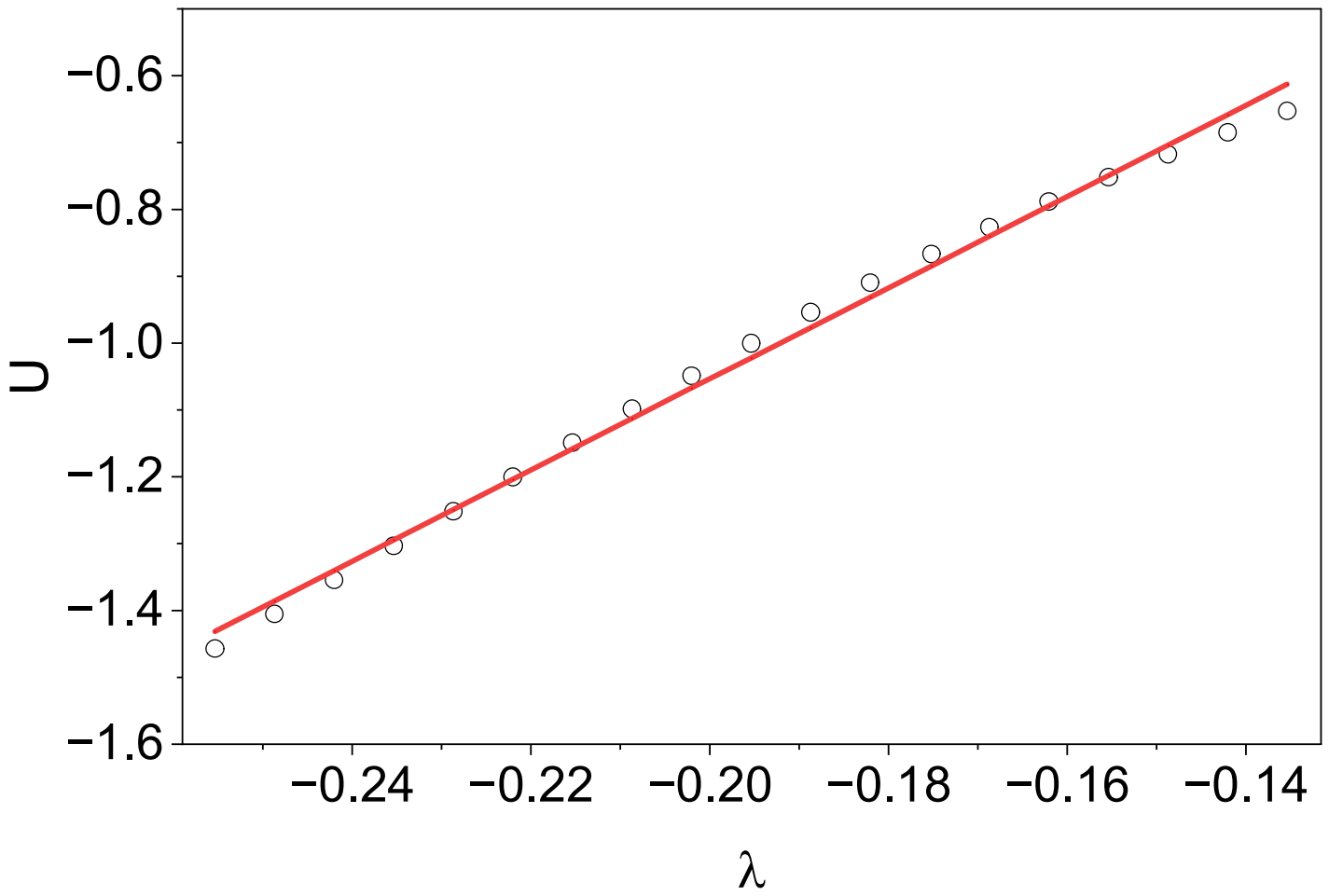

Fig. S5 The U vs. $\lambda$ plot for Al in the Ni-5Re side of the diffusion couple produced by coupling Ni-5Re and Ni-5Al at 1200 ºC.

## S8 Optmization parameters established in Ni-Al-X systems

Table S3 Optmization parameters for calculation of tracer diffusion coefficients in the Ni-Al-X systems covering the composition ranges of the ternary diffusion profiles

| $D_{Ni}^{*}$ (Ni-Cr-Al) | Parameter | 1100 °C | 1150 °C | 1200 °C |
|---|---|---|---|---|
| $D_{Ni}^{*}$ | $\theta_{0}^{Ni}$ | 2.3 | 3.2 | 4.0 |
| | $\theta_{Cr}^{1,Ni}$ | 0.0026 | -0.009 | -0.0005 |
| | $\theta_{Al}^{1,Ni}$ | -0.019 | 0.01 | -0.0002 |
| | $\theta_{Cr,Al}^{2,Ni}$ | 0.5 | -0.07 | 0.02 |
| $D_{Cr}^{*}$ | $\theta_{0}^{Cr}$ | 2.8 | 3.7 | 4.5 |
| | $\theta_{Cr}^{1,Cr}$ | -3.9 | -10.7 | -7.5 |
| | $\theta_{Al}^{1,Cr}$ | -4.6 | -9.8 | -5.8 |
| | $\theta_{Cr,Al}^{2,Cr}$ | -2.4 | -8.7 | -6.9 |
| $D_{Al}^{*}$ | $\theta_{0}^{Al}$ | 3.8 | 4.6 | 5.3 |
| | $\theta_{Cr}^{1,Al}$ | -4.2 | -7.2 | -1.1 |
| | $\theta_{Al}^{1,Al}$ | -5.9 | -6.5 | -0.8 |
| | $\theta_{Cr,Al}^{2,Al}$ | -4.4 | -5.7 | -0.8 |

(a)

| $D_{Ni}^{*}$ (Ni-Mo-Al) | Parameter | 1100 °C | 1150 °C | 1200 °C |
|---|---|---|---|---|
| $D_{Ni}^{*}$ | $\theta_{0}^{Ni}$ | 2.3 | 3.2 | 4.0 |
| | $\theta_{Mo}^{1,Ni}$ | 0.03 | 0.0008 | 0.1 |
| | $\theta_{Al}^{1,Ni}$ | -0.06 | 0.0008 | 0.5 |
| | $\theta_{Mo,Al}^{2,Ni}$ | 0.5 | -0.05 | 0.9 |
| $D_{Mo}^{*}$ | $\theta_{0}^{Mo}$ | 2.1 | 3.0 | 3.3 |
| | $\theta_{Cr}^{1,Mo}$ | -7.3 | -6.1 | 0.4 |
| | $\theta_{Al}^{1,Mo}$ | -6.2 | -4.1 | 0.7 |
| | $\theta_{Mo,Al}^{2,Mo}$ | -1.8 | -2.2 | 0.9 |
| $D_{Al}^{*}$ | $\theta_{0}^{Al}$ | 3.9 | 4.7 | 4.8 |
| | $\theta_{Mo}^{1,Al}$ | -13.8 | -10.2 | -0.09 |
| | $\theta_{Al}^{1,Al}$ | -11.7 | -7.0 | 0.4 |
| | $\theta_{Mo,,Al}^{2,Al}$ | -12.0 | -9.6 | 0.6 |

(b)

| $D_{Ni}^{*}$ (Ni-Ta-Al) | Parameter | 1100 °C | 1150 °C | 1200 °C |
|---|---|---|---|---|
| $D_{Ni}^{*}$ | $\theta_{0}^{Ni}$ | 2.3 | 3.2 | 4.0 |
| | $\theta_{Ta}^{1,Ni}$ | -0.006 | 0.009 | -0.003 |
| | $\theta_{Al}^{1,Ni}$ | -0.01 | -0.01 | 0.008 |
| | $\theta_{Ta,Al}^{2,Ni}$ | 0.6 | 0.0007 | -0.1 |
| $D_{Ta}^{*}$ | $\theta_{0}^{Ta}$ | 3.2 | 4.0 | 4.7 |
| | $\theta_{Ta}^{1,Ta}$ | -8.8 | -10.6 | -10.1 |
| | $\theta_{Al}^{1,Ta}$ | -5.7 | -10.0 | -8.5 |
| | $\theta_{Ta,Al}^{2,Ta}$ | -0.5 | -9.6 | -9.8 |
| $D_{Al}^{*}$ | $\theta_{0}^{Al}$ | 3.9 | 4.5 | 5.3 |
| | $\theta_{Ta}^{1,Al}$ | -14.0 | 2.7 | 2.2 |
| | $\theta_{Al}^{1,Al}$ | -11.5 | 2.6 | 2.0 |
| | $\theta_{Ta,Al}^{2,Al}$ | -12.4 | 2.1 | 2.0 |

(c)

| $D_{Ni}^{*}$ (Ni-W-Al) | Parameter | 1100 °C | 1150 °C | 1200 °C |
|---|---|---|---|---|
| $D_{Ni}^{*}$ | $\theta_{0}^{Ni}$ | 2.3 | 3.2 | 4.0 |
| | $\theta_{W}^{1,Ni}$ | 0.004 | 0.05 | 0.01 |
| | $\theta_{Al}^{1,Ni}$ | -0.04 | -0.2 | -0.06 |
| | $\theta_{W,Al}^{2,Ni}$ | 0.3 | 0.6 | 0.3 |
| $D_{W}^{*}$ | $\theta_{0}^{W}$ | 0.7 | 1.78 | 2.6 |
| | $\theta_{W}^{1,W}$ | -8.4 | -12.0 | -14.5 |
| | $\theta_{Al}^{1,W}$ | -5.7 | -10.4 | -10.8 |
| | $\theta_{W,Al}^{2,W}$ | -0.5 | -3.6 | -5.5 |
| $D_{Al}^{*}$ | $\theta_{0}^{Al}$ | 3.9 | 4.7 | 5.5 |
| | $\theta_{W}^{1,Al}$ | -14.0 | -12.5 | -14.3 |
| | $\theta_{Al}^{1,Al}$ | -11.5 | -11.2 | -10.1 |
| | $\theta_{W,Al}^{2,Al}$ | -12.4 | -12.4 | -12.2 |

(d)

| $D_{Ni}^{*}$ (Ni-W-Al) | Parameter | 1150 °C | 1200 °C | 1250 °C |
|---|---|---|---|---|
| $D_{Ni}^{*}$ | $\theta_{0}^{Ni}$ | 3.2 | 3.9 | 4.7 |
| | $\theta_{Re}^{1,Ni}$ | -0.01 | 0.5 | 0.6 |
| | $\theta_{Al}^{1,Ni}$ | -0.02 | 1.1 | 1.2 |
| | $\theta_{Re,Al}^{2,Ni}$ | 0.4 | 0.9 | 0.7 |
| $D_{Re}^{*}$ | $\theta_{0}^{Re}$ | -0.6 | 0.1 | 1.1 |
| | $\theta_{Cr}^{1,Re}$ | 11.1 | 11.1 | 10.5 |
| | $\theta_{Al}^{1,Re}$ | -10.2 | -7.3 | -8.6 |
| | $\theta_{Mo,Al}^{2,Re}$ | 0.3 | 1.5 | 4.8 |
| $D_{Al}^{*}$ | $\theta_{0}^{Al}$ | 4.1 | 4.7 | 5.5 |
| | $\theta_{Mo}^{1,Al}$ | 0.3 | 0.4 | 0.4 |
| | $\theta_{Al}^{1,Al}$ | 0.8 | 1.3 | 1.3 |
| | $\theta_{Re,,Al}^{2,Al}$ | 0.5 | 0.3 | 0.3 |

(e)